\documentclass{article}
\usepackage{ijcai26}

\usepackage{times}
\usepackage{soul}
\usepackage{url}
\usepackage[hidelinks]{hyperref}
\usepackage[utf8]{inputenc}
\usepackage[small]{caption}
\usepackage{graphicx}
\usepackage{amsmath}
\usepackage{amsthm}
\usepackage{booktabs}
\usepackage{algorithm}
\usepackage{algorithmic}
\usepackage[switch]{lineno}
\usepackage{enumitem}
\usepackage{booktabs}
\usepackage{amssymb}
\usepackage{subcaption}
\usepackage{xcolor}
\usepackage{placeins}

\title{Prism: Spectral Parameter Sharing for Multi-Agent Reinforcement Learning}

\author{
Kyungbeom Kim$^1$
\and
Seungwon Oh$^1$\And
Kyung-Joong Kim$^1$\\
\affiliations
$^1$Gwangju Institute of Science and Technology (GIST)\\
\emails
\{kyungbeom8, sw980907\}@gm.gist.ac.kr,
kjkim@gist.ac.kr
}

\begin{document}

\maketitle

\begin{abstract}
Parameter sharing is a key strategy in multi-agent reinforcement learning (MARL) for improving scalability, yet conventional fully shared architectures often collapse into homogeneous behaviors. Recent methods introduce diversity through clustering, pruning, or masking, but typically compromise resource efficiency. We propose \textbf{Prism}, a parameter sharing framework that induces inter-agent diversity by representing shared networks in the spectral domain via singular value decomposition (SVD). All agents share the singular vector directions while learning distinct spectral masks on singular values. This mechanism encourages inter-agent diversity and preserves scalability. Extensive experiments on both homogeneous (LBF, SMACv2) and heterogeneous (MaMuJoCo) benchmarks show that \textbf{Prism} achieves competitive performance with superior resource efficiency.
\end{abstract}

%%%%%%%%%%%%%%%%%%%%%%%%%%%%%%%%%%%%%%%%%%%%%%%%%%%%%%%%%%%%%%%%%%%%%%%%
\section{Introduction}
Multi-Agent Reinforcement Learning (MARL) has emerged as a powerful paradigm for tackling complex real-world problems such as traffic signal control~\cite{MDPI_2023_traffic_control}, robotics~\cite{AAAI_2025_robotics}, and autonomous driving~\cite{IROS_2022_autonomous_driving}. These applications often require learning coordinated behaviors among tens to hundreds of agents, making scalability a central challenge.
While Centralized Training with Decentralized Execution (CTDE) ~\cite{NIPS2017_68a97503} alleviates non-stationarity and credit assignment problems by leveraging global information during training, it does not fundamentally resolve the scalability problem, as model size, computational cost, and memory usage still grow rapidly with the number of agents. To address this issue, parameter sharing has been widely adopted, enabling all agents to learn using a shared set of parameters and significantly reducing training and memory overhead. However, this efficiency comes at the cost of expressiveness: a single shared policy often fails to capture agent heterogeneity, limiting its ability to learn optimal behaviors ~\cite{NEURIPS2020_7967cc8e,pmlr-v162-fu22d}. Although incorporating agent identities has been proposed to induce policy heterogeneity, such approaches remain insufficient in tasks that require substantial heterogeneity across agents and may even hinder learning in practice ~\cite{pmlr-v139-christianos21a,tessera2025hypermarl}. 

To overcome these limitations, recent studies have explored partial parameter sharing frameworks that mitigate excessive policy homogeneity by either clustering agents based on early data, or inducing agent-specific diversity through node- or edge-level pruning and masking within a shared network ~\cite{pmlr-v139-christianos21a,li2024adaptive,kim2023parameter,NEURIPS2024_274d0146}. However, clustering-based methods are highly sensitive to the initial data distribution, posing significant challenges in non-stationary environments. Meanwhile, pruning- and masking-based methods require maintaining explicit representations of nodes and edges, leading to a rapid decline in resource efficiency as the number of agents increases. These limitations highlight a persistent trade-off between scalability and policy heterogeneity in parameter sharing MARL frameworks.
\begin{figure}[t]
    \centering
    \includegraphics[width=\columnwidth]{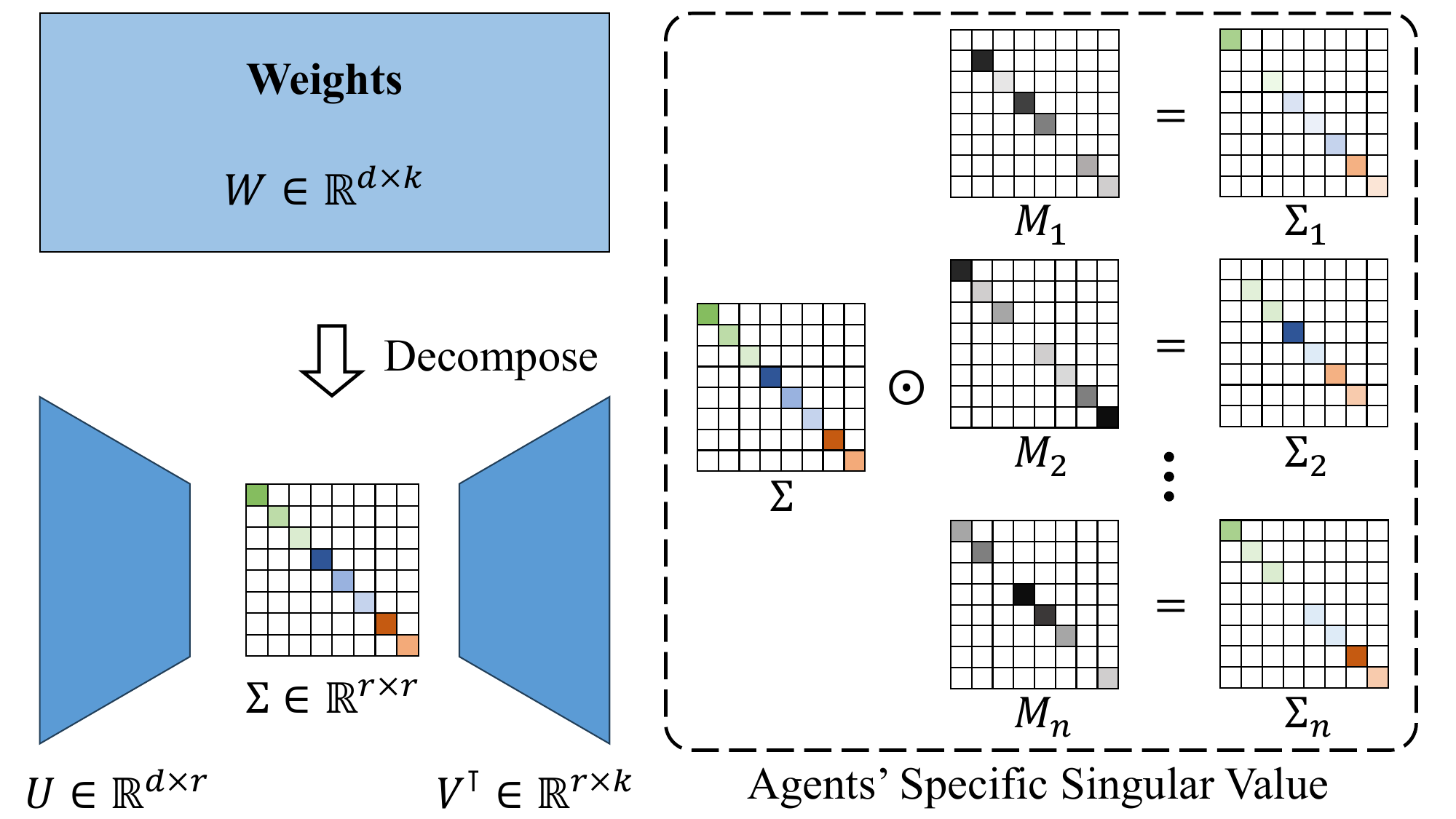}
    \caption{
        \textbf{Overview of Prism framework.}
        We decompose the shared weights $W\in\mathbb{R}^{d\times k}$ into $W=U\Sigma V^\top$ and learn them in the SVD-parameterized space. The left/right singular vectors $U$ and $V$ are shared among agents, whereas agent $i$ applies a learnable spectral mask to the singular values to obtain an agent-specific spectrum $\Sigma_i$, enabling diversity while preserving parameter sharing.
    }
    \label{fig:overview}
\end{figure}

%%%%%%%%%%%%%%%%%%%%%%%%%%%%%%%%%%%%%%%%%%%%%%%%
% In this paper, we introduce \textbf{Prism}, a parameter-sharing framework that draws on the idea of spectral diversity to induce inter-agent diversity in the spectral space rather than in the weight space. Concretely, we employ learnable masks in the spectral space to adaptively allocate each agent an appropriate subspace - in a manner similar to how a prism decomposes a single ray of light into diverse spectral components. 
% This design allows our method to address complex MARL tasks without being restricted by the initial data distribution, 
% while also requiring fewer resources compared to previous node- and edge-level masking approaches.Through extensive experiments, we demonstrate that spectral diversity enables the simultaneous achievement of both performance and resource efficiency.
In this paper, we propose \textbf{Prism}, a novel parameter sharing that operates in the spectral space rather than the weight space.
Specifically, we employ learnable masks in the spectral space to adaptively allocate each agent an appropriate subspace. This design allows our method to address complex MARL tasks without being restricted by the initial data distribution, 
while also requiring fewer resources compared to previous node- and edge-level masking approaches. Through extensive experiments, we demonstrate that spectral parameter sharing allows the simultaneous achievement of policy heterogeneity and memory efficiency across diverse MARL benchmarks. 

%%%%%%%%%%%%%%%%%%%%%%%%%%%%%%%%%%%%%%%%%%%%%%%
The main contributions are as follows:
\begin{enumerate}[topsep=0.5em]
    \item We propose a spectral parameter sharing framework that induces agent differentiation by allocating distinct spectral subspaces within a shared network.
    \item We mitigate the inherent trade-off between scalability and policy heterogeneity by inducing agent differentiation solely in the spectral space, thereby avoiding redundant parameters and substantially improving resource efficiency.
    \item We empirically validate the effectiveness of the proposed approach through extensive experiments on diverse MARL benchmarks, demonstrating competitive or superior performance compared to existing methods while significantly reducing memory overhead.
\end{enumerate}

\section{Background}

% \subsection{Dec-POMDP}

We formulate the multi-agent reinforcement learning problem as a
Decentralized Partially Observable Markov Decision Process (Dec-POMDP)~\cite{oliehoek2016concise}. 
A Dec-POMDP is defined by a tuple 
\[
(\mathcal{S}, \{\mathcal{A}_i\}_{i=1}^n, \{\mathcal{O}_i\}_{i=1}^n, P, R, \mathcal{N}, \gamma),
\]
where \(\mathcal{S}\) is the global state space, 
\(\mathcal{A}_i\) and \(\mathcal{O}_i\) denote the action and observation spaces of agent \(i\), respectively, 
\(P: \mathcal{S} \times \mathcal{A}_1 \times \dots \times \mathcal{A}_n \rightarrow \Delta(\mathcal{S})\) is the state transition function, 
\(R: \mathcal{S} \times \mathcal{A}_1 \times \dots \times \mathcal{A}_n \rightarrow \mathbb{R}\) is the shared reward function, 
\(\mathcal{N} = \{1, \dots, n\}\) is the set of agents, 
and \(\gamma \in [0,1)\) is the discount factor.  

At each timestep, the environment is in a state \(s \in \mathcal{S}\). Each agent \(i \in \mathcal{N}\) receives a local observation \(o_i \in \mathcal{O}_i\) and selects an action \(a_i \in \mathcal{A}_i\) according to its policy \(\pi_i(a_i \mid o_i)\). In fully observable settings (a special case of the Dec-POMDP framework), the policy can equivalently take the global state \(s\) as input, i.e., \(\pi_i(a_i \mid s)\). The joint action \(\mathbf{a} = (a_1, \dots, a_n)\) induces a transition to the next state \(s' \sim P(\cdot \mid s, \mathbf{a})\), and a shared reward \(r = R(s, \mathbf{a})\) is obtained. The objective is to learn decentralized policies \(\{\pi_i\}_{i=1}^n\) that maximize the expected cumulative discounted reward:
\[
J(\pi) = \mathbb{E}\left[\sum_{t=0}^{\infty} \gamma^t R(s_t, \mathbf{a}_t)\right],
\]
where the expectation is taken over the trajectories induced by the joint policy \(\pi = \{\pi_i\}_{i=1}^n\) and the transition dynamics.

\section{Spectral Parameter Sharing Formulation}

In multi-agent reinforcement learning, each agent’s policy weight matrix can be viewed as a mixture of shared basis components.
Let $\{\mathbf{B}_1,\dots,\mathbf{B}_r\}$ be shared basis components (with arbitrary scales) where $r$ controls representational capacity. Specifically, each agent is associated with a coefficient $\boldsymbol{\alpha}^{(i)} \in \mathbb{R}^r$, and its weight matrix can be approximated as
\[
\mathbf{W}^{(i)} \approx \sum_{j=1}^{r} \alpha^{(i)}_j\mathbf{B}_j,
\qquad \boldsymbol{\alpha}^{(i)} = (\alpha^{(i)}_1, \dots, \alpha^{(i)}_r).
\]
The coefficients $\boldsymbol{\alpha}^{(i)}$ determine how each agent utilizes the shared bases, effectively balancing between common and agent-specific representations.

\paragraph{\textbf{Orthogonality to Reduce Interference and Redundancy.}}
The overlap between two matrices can be measured by the Frobenius inner product 
$\langle \mathbf{A},\mathbf{B}\rangle_F := \mathrm{tr}(\mathbf{A}^\top \mathbf{B})$. 
When basis components ${\mathbf{B}_j}$ are correlated (i.e., $\langle \mathbf{B}_p,\mathbf{B}_q\rangle_F \neq 0$ for $p \neq q$), their representations become entangled through cross-terms $\langle \mathbf{B}_p,\mathbf{B}_q\rangle_F$. 
These cross-terms cause the coefficient of one component to depend on others, 
so that updating or scaling a single basis element can unintentionally alter neighboring ones. 
This dependence results in \emph{interference} between basis components 
and leads to \emph{redundancy} by allowing multiple bases to encode overlapping information. 
Enforcing orthogonality, 
\[
\langle \mathbf{B}_p,\mathbf{B}_q\rangle_F = 0 \quad (p\neq q),
\]
eliminates such cross-terms and ensures that each basis contributes independently. 
Under this condition, the coefficient of component $j$ admits a clean projection form,
\[
\alpha^{(i)}_j = 
\frac{\langle \mathbf{W}^{(i)}, \mathbf{B}_j \rangle_F}{\|\mathbf{B}_j\|_F^2},
\]
where each $\alpha^{(i)}_j$ reflects the isolated contribution of $\mathbf{B}_j$ 
without being affected by others. 
Consequently, orthogonality prevents interference among basis components 
and avoids redundant representations, 
resulting in a more efficient and interpretable parameterization.

\paragraph{\textbf{Singular Value Decomposition as an Orthogonal Basis.}}
To construct an orthogonal set of rank-1 bases $\{\mathbf{B}_j\}$, 
we introduce orthonormal matrices 
$\mathbf{U}\!\in\!\mathbb{R}^{d\times r}$ and $\mathbf{V}\!\in\!\mathbb{R}^{k\times r}$ 
satisfying $\mathbf{U}^\top\mathbf{U}=\mathbf{I}_r$ and $\mathbf{V}^\top\mathbf{V}=\mathbf{I}_r$, 
and define
\[
\mathbf{B}_j := \mathbf{u}_j \mathbf{v}_j^\top, \quad j=1,\dots,r,
\]
where $\mathbf{u}_j$ and $\mathbf{v}_j$ are the $j$-th columns of $\mathbf{U}$ and $\mathbf{V}$. 
Each $\mathbf{B}_j$ corresponds to a single pair of left and right singular vectors, 
forming a mutually orthogonal \emph{shared spectral basis} across agents. 
This basis acts as a common parameterization frame in which each agent’s weights are represented as
\[
\mathbf{W}^{(i)} \;\approx\; 
\mathbf{U}\,\mathrm{diag}\!\big(\boldsymbol{\sigma}\odot \boldsymbol{\alpha}^{(i)}\big)\,\mathbf{V}^\top,
\]
with $\boldsymbol{\sigma}\!\in\!\mathbb{R}^r_{\ge 0}$ denoting the shared singular spectrum and $\boldsymbol{\alpha}^{(i)}$ an agent-specific scaling vector. 
Such factorization provides an orthogonal coordinate structure for parameter sharing, 
allowing agents to specialize by selectively scaling different spectral modes.

\section{Method}
In this section, we propose a novel parameter sharing method that models weights through an SVD-parameterized low-rank factorization.
Each agent applies a learnable spectral mask on the singular values to enable adaptive and diverse spectral spaces.
An overview of the proposed method is illustrated in Figure \ref{fig:overview}.
% Building upon the theoretical formulation in the previous section, we first present our empirical motivation to verify whether redundancy in parameter sharing arises from the low intrinsic dimensionality of multi-agent policies within high capacity shared networks.
% This observation further supports the redundancy identified in parameter-sharing architectures and highlights the need for a more principled approach to exploiting shared representations.
% We hypothesize that such redundancy can be more effectively utilized through spectral factorization, enabling each agent to occupy a distinct representational subspace.
% To this end, we propose \textbf{Prism}, a novel parameter-sharing framework that models shared weights through SVD-based parameterization.
% Each agent applies a learnable spectral mask to the singular values to disentangle shared and agent-specific components, with additional regularization promoting diversity and orthogonality.
% An overview of the proposed framework is illustrated in Figure~\ref{fig:overview}.

\subsection{SVD Decomposed Weight}
% To address this redundancy, we introduce a spectral parameterization of shared weights based on singular value decomposition (SVD). As illustrated in Figure ~\ref{fig:overview}. (a).
We parameterize the weight matrix of the shared network using an SVD factorization, where
$\mathbf{W} \in \mathbb{R}^{d \times k}$ is decomposed as
\[
\mathbf{W}_\theta = \mathbf{U} \, \mathrm{diag}(\mathbf{s}) \, \mathbf{V^\top},
\]
where $\mathbf{U} \in \mathbb{R}^{d \times r}$ and $\mathbf{V} \in \mathbb{R}^{k \times r}$ denote the left and right singular vector matrices, and 
$\mathbf{s} \in \mathbb{R}^{r}$ corresponds to the singular values, with $r = \min(d, k)$. 
All components $(\mathbf{U}, \mathbf{s}, \mathbf{V})$ are treated as learnable parameters, enabling end-to-end training in the SVD-parameterized space.
This factorization provides the foundation of inducing diversity in the spectral space, which will be introduced through spectral space masking in the next subsection.

\subsection{Spectral Space Masking}
To induce agent-specific diversity within the SVD-parameterized shared network, we introduce the spectral space masking.
Each agent $i$ is associated with a learnable mask $\mathbf{m}_i \in \mathbb{R}^r$ applied to the singular values. 
In this design, the singular vector matrices $\mathbf{U}$ and $\mathbf{V}$ are shared across all agents to maintain a common basis, 
while diversity is introduced exclusively through the singular values $\mathbf{s}$, which are adaptively differentiated for each agent via learnable spectral masks. 
Formally, the agent-specific spectral values are defined as

\[
\mathbf{s}_i = \mathrm{concat}(\mathbf{s}_{\text{common}}, (\mathbf{s}_{\text{separate}} \odot \mathbf{m}_i)),
\]
where $\mathbf{s}_{\text{common}}$ denotes the shared spectral components, $\mathbf{s}_{\text{separate}}$ represents the agent-specific part scaled by the mask $\mathbf{m}_i$, and $\odot$ denotes the Hadamard product (see Figure \ref{fig:overview}). 
The mask is parameterized as
\[
\mathbf{m}_i = \mathrm{ReLU}\!\left( \mathbf{s}_{\text{norm}} - \sigma\!\left(\mathbf{t}_\Psi^{(i)}\right) \right),
\]
where $\mathbf{t}_\Psi^{(i)} \in \mathbb{R}^r$ denotes learnable threshold parameters for agent $i$, 
$\sigma(\cdot)$ is the sigmoid activation, and 
\(\mathbf{s}_{\text{norm}} = \mathbf{s}_{\text{separate}} / \max(\mathbf{s}_{\text{separate}})\).

This formulation allows flexible decomposition of the spectral space into common and agent-specific components, with their relative proportion controlled by a hyperparameter.
As a result, agents can be adaptively allocated to distinct spectral subspaces, enabling agent differentiation while preserving the benefits of parameter sharing.

\subsection{Diversity and Orthogonal Regularization}
While spectral space masking enables agent-specific parameterization, additional regularization is required to ensure sufficient inter-agent diversity and stable training. 
We introduce two complementary regularization terms: diversity regularization and orthogonal regularization. 

\paragraph{\textbf{Diversity Regularization.}}
To explicitly encourage difference among agents in the spectral space, we penalize similarity between their masks. 
During training, the continuous masks $\mathbf{m}_i$ are used to maintain differentiability, 
while a binary mask $\tilde{\mathbf{m}}_i$ serves as a conceptual approximation for the active spectral regions in the computation of the diversity loss. 
Formally, the binary mask is defined as
\[
\tilde{\mathbf{m}}_i = \mathbb{I}[\mathbf{m}_i > 0],
\]
where $\mathbb{I}[\cdot]$ denotes the indicator function. 
The diversity regularization term is then expressed as
\[
J^{\text{div}}(s) =
\sum_{i=1,\ldots,n} \sum_{j=1, j \neq i}^{n}
\left\|
\mathbf{s}_{\text{separate}} \odot
\left( \tilde{\mathbf{m}}_i - \tilde{\mathbf{m}}_j \right)
\right\|_1,
\]
where $\odot$ denotes the Hadamard product. 
Since the binary masking operation is non-differentiable, we employ the Straight-Through Estimator (STE) \cite{bengio2013estimating} to approximate the gradients during backpropagation. 
This objective guides agents to attend to distinct spectral components, thereby mitigating redundancy and promoting heterogeneous representations.

\paragraph{\textbf{Orthogonal Regularization.}}
To preserve the stability of the SVD decomposition, we additionally regularize the singular vector matrices $\mathbf{U}$ and $\mathbf{V}$ towards orthogonality: 
\[
\mathcal{L}_{\text{ortho}} = \left\| \mathbf{U}^\top \mathbf{U} - \mathbf{I}_r \right\|_F^2 
+ \left\| \mathbf{V}\mathbf{V}^\top - \mathbf{I}_r \right\|_F^2,
\]
where $\|\cdot\|_F$ denotes the Frobenius norm and $\mathbf{I}_r$ is the identity matrix of size $r$. 
This regularization mitigates the collapse of singular vectors and stabilizes end-to-end training under the SVD factorization.
% We summarize the overall training procedure of the \textbf{Prism} framework in Algorithm~\ref{alg:prism} which integrates spectral parameterization, spectral subspace masking, and the proposed regularization terms within a standard centralized training and decentralized execution paradigm.

\begin{figure*}[t]
    \centering
    
    % 상단 legend 이미지
    \includegraphics[width=0.4\textwidth]{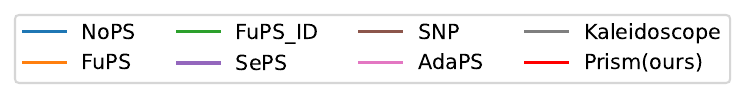}
    \vspace{0.5em} % 살짝 간격 조절
    
    % 하단 3개의 이미지 (각자 caption 포함)
    \begin{subfigure}[t]{0.24\textwidth}
        \centering
        \includegraphics[width=\textwidth]{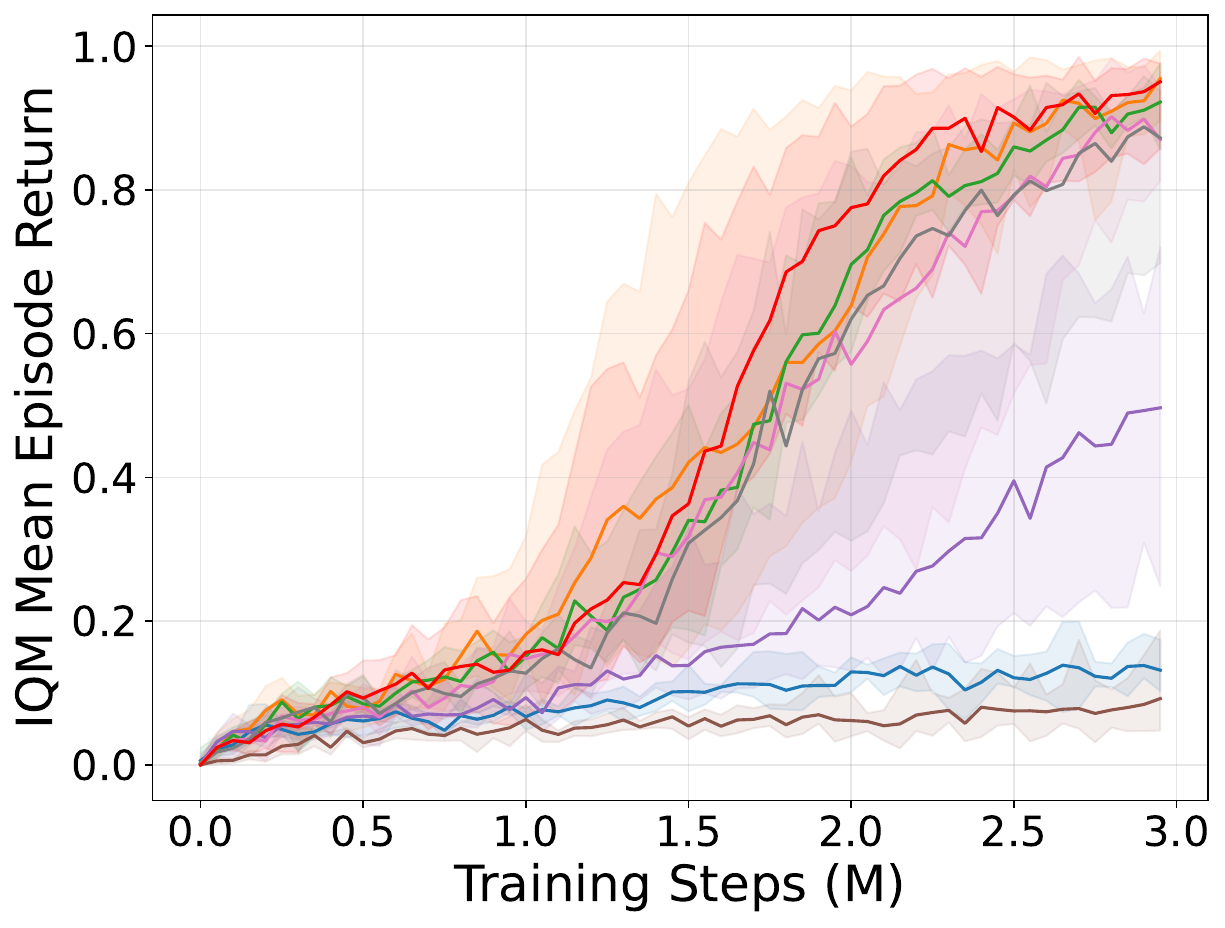}
        \captionsetup{skip=0pt}
        \caption{lbf-10x10-3p-3f}
        \label{fig:lbf}
    \end{subfigure}%
    \begin{subfigure}[t]{0.24\textwidth}
        \centering
        \includegraphics[width=\textwidth]{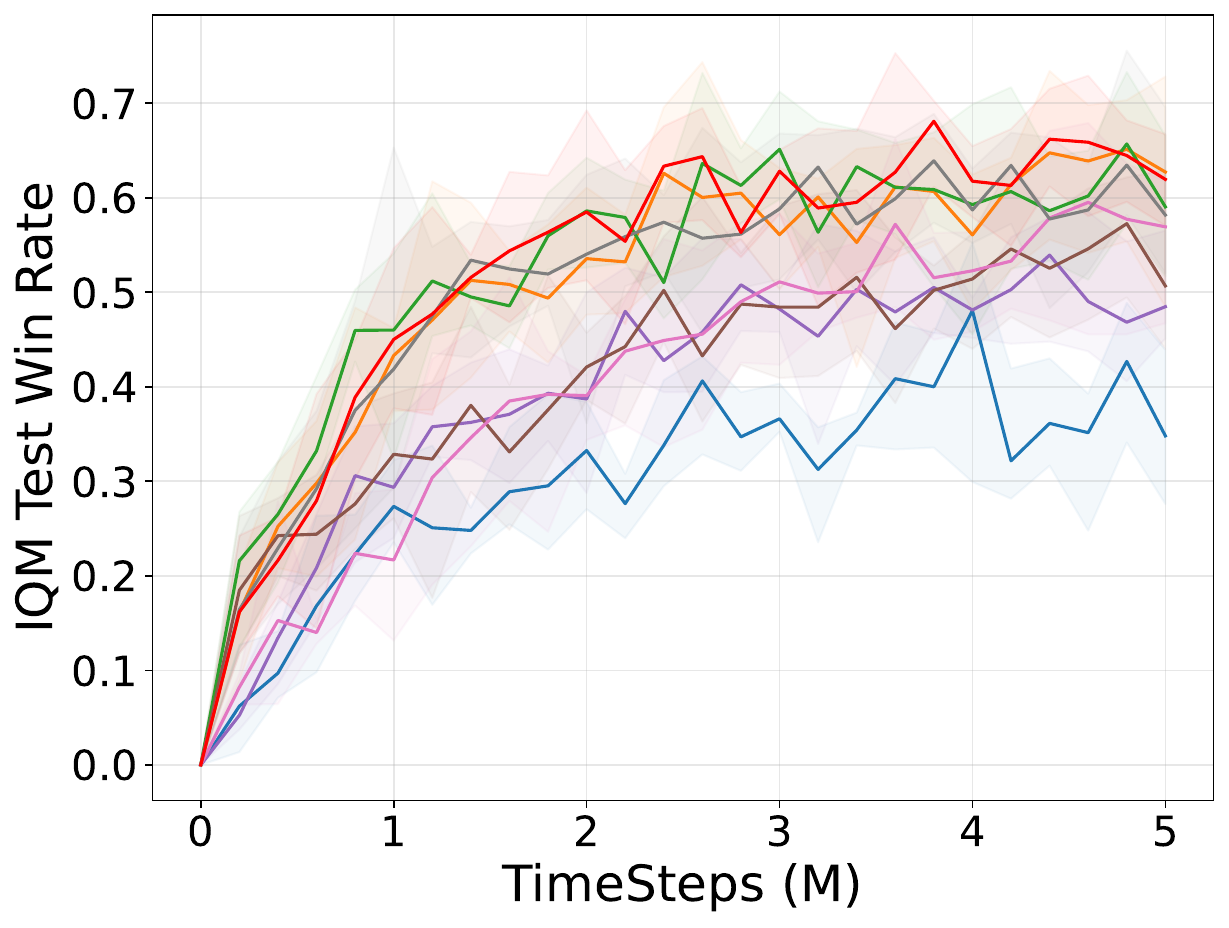}
        \captionsetup{skip=0pt}
        \caption{Terran\_5\_vs\_5}
        \label{fig:smacv2_terran}
    \end{subfigure}%
    \hspace{0.5em}
    \begin{subfigure}[t]{0.24\textwidth}
        \centering
        \includegraphics[width=\textwidth]{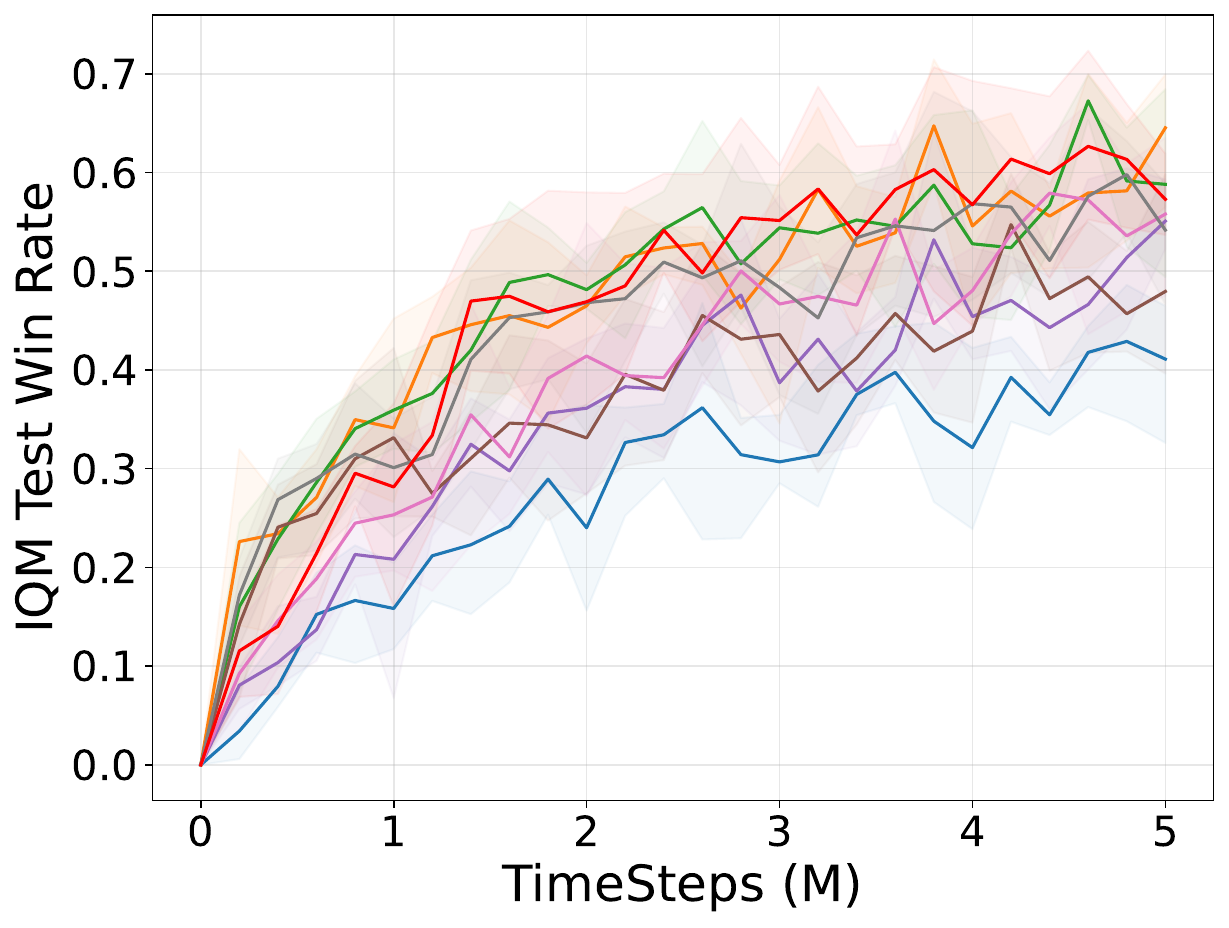}
        \captionsetup{skip=0pt}
        \caption{Protoss\_5\_vs\_5}
        \label{fig:smacv2_protoss}
    \end{subfigure}%
    \hspace{0.5em}
    \begin{subfigure}[t]{0.24\textwidth}
        \centering
        \includegraphics[width=\textwidth]{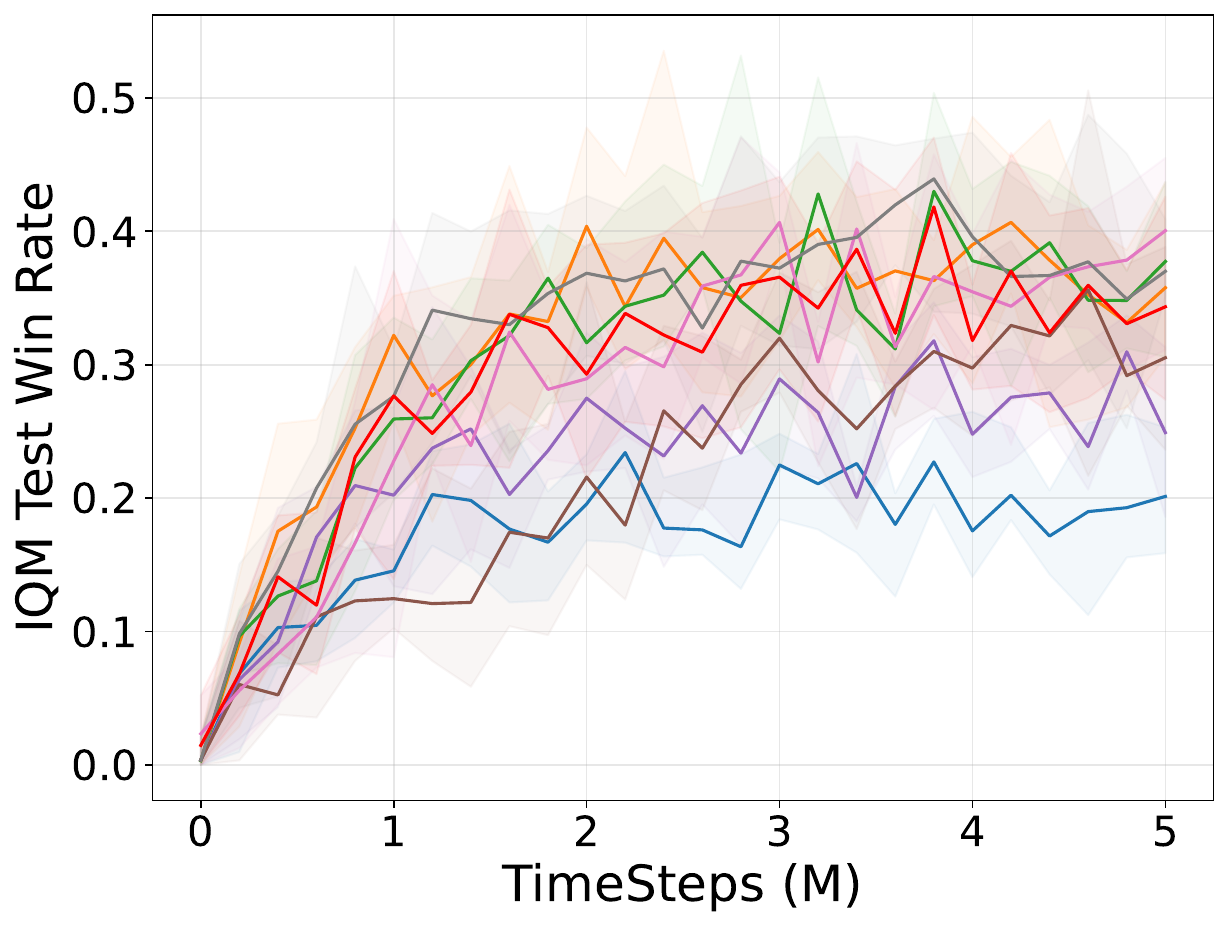}
        \captionsetup{skip=0pt}
        \caption{Zerg\_5\_vs\_5}
        \label{fig:smacv2_zerg}
    \end{subfigure}
    % 전체 figure 캡션
    \captionsetup{skip=1pt}
    \caption{Performance evaluation on homogeneous environments (LBF, SMACv2).}
    \label{fig:smacv2_performance}
\end{figure*}

\begin{figure*}[t]
    \centering % <--- [핵심] 전체 그림을 가운데로 정렬
    
    % --- 첫 번째 줄 (a, b, c) ---
    \begin{subfigure}[t]{0.25\textwidth}
        \centering
        \includegraphics[width=\textwidth]{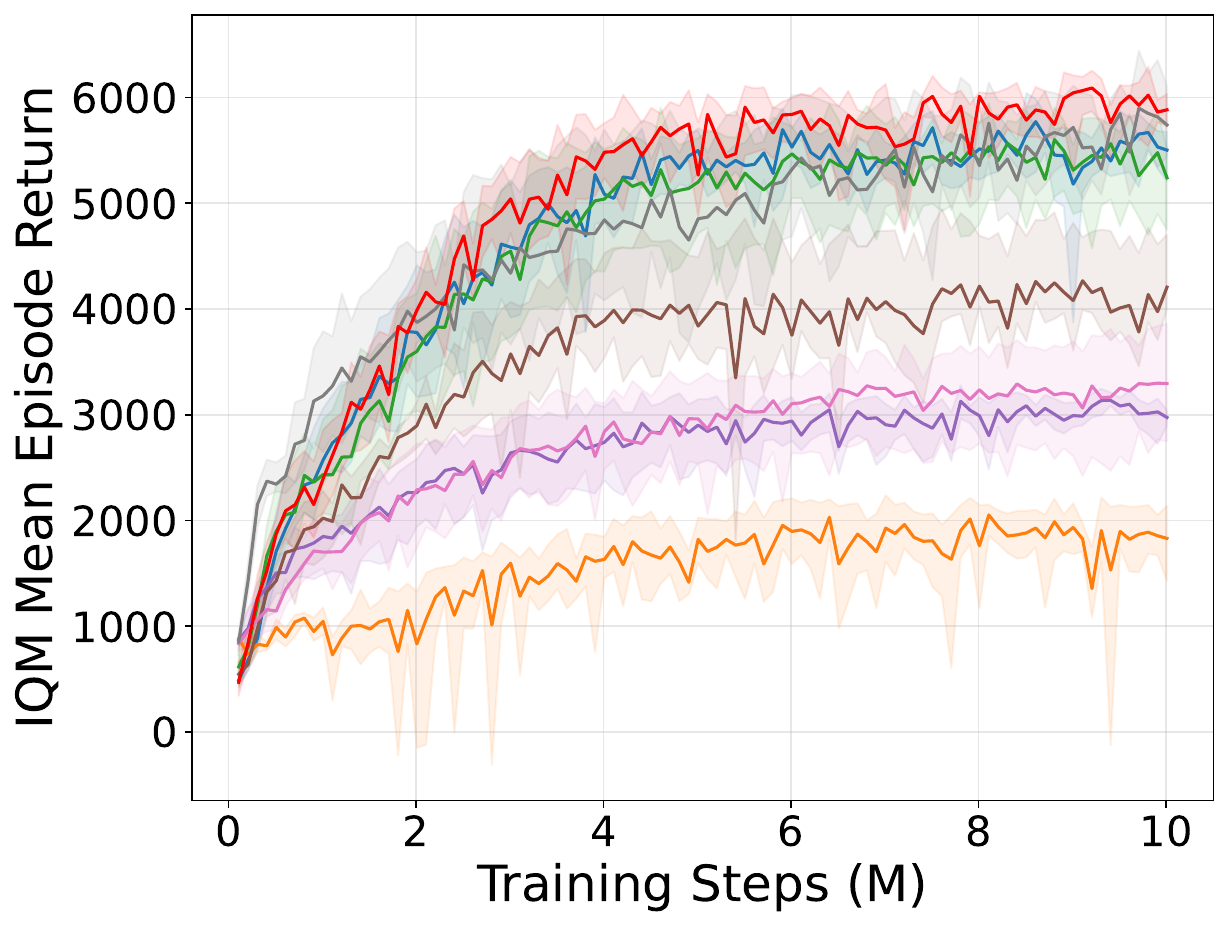}
        \captionsetup{skip=0pt}
        \caption{Ant-v2-4x2}
    \end{subfigure}%
    \hspace{0.5em}
    \begin{subfigure}[t]{0.25\textwidth}
        \centering
        \includegraphics[width=\textwidth]{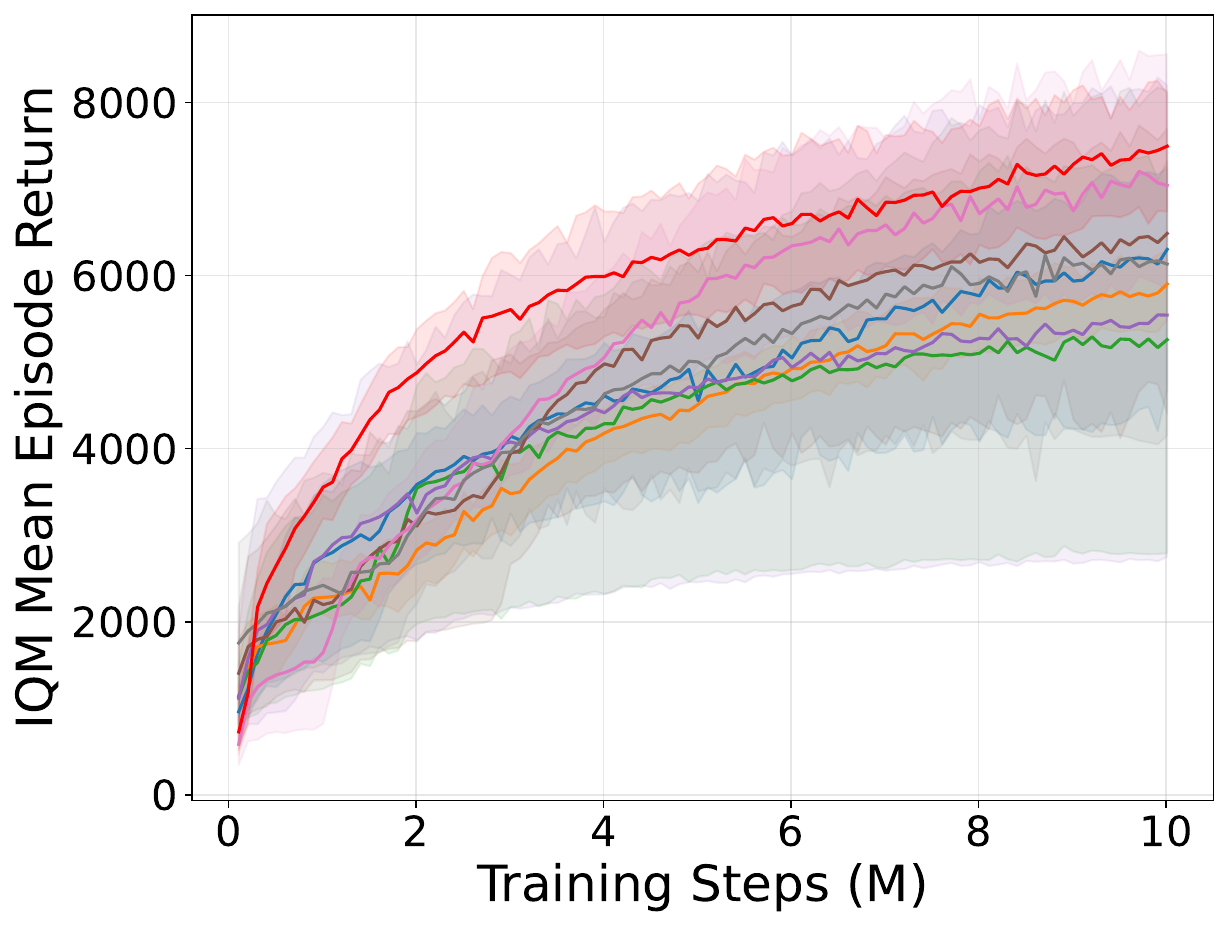}
        \captionsetup{skip=0pt}
        \caption{HalfCheetah-v2-2x3}
    \end{subfigure}%
    \hspace{0.5em}
    \begin{subfigure}[t]{0.25\textwidth}
        \centering
        \includegraphics[width=\textwidth]{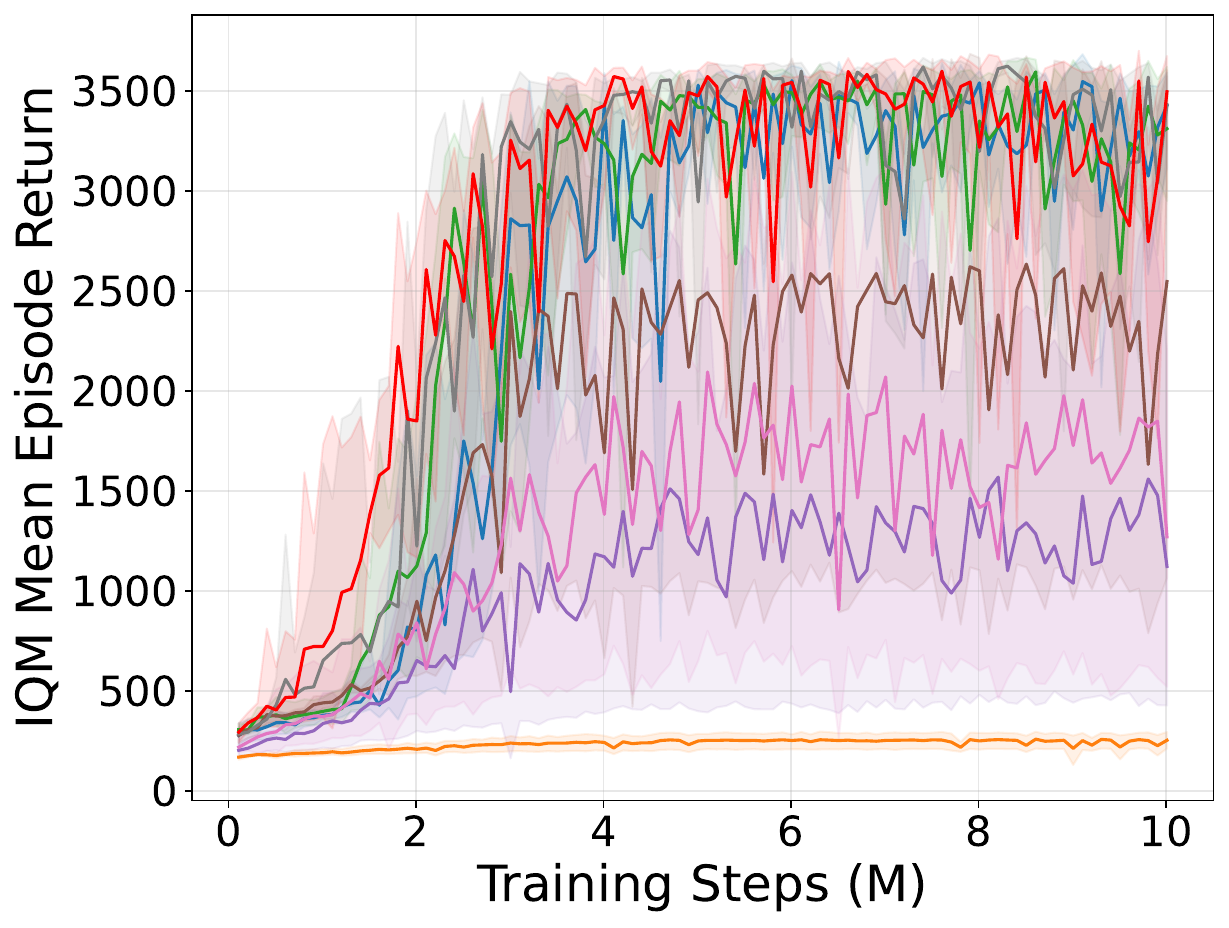}
        \captionsetup{skip=0pt}
        \caption{Hopper-v2-3x1}
    \end{subfigure}
    
    % \vspace{1em} % 첫 번째 줄과 두 번째 줄 사이 간격
    % \newline % <--- [핵심] 여기서 줄을 바꿈 (3개 배치 후 다음 줄로)

    % --- 두 번째 줄 (d, e, f) ---
    \begin{subfigure}[t]{0.25\textwidth}
        \centering
        \includegraphics[width=\textwidth]{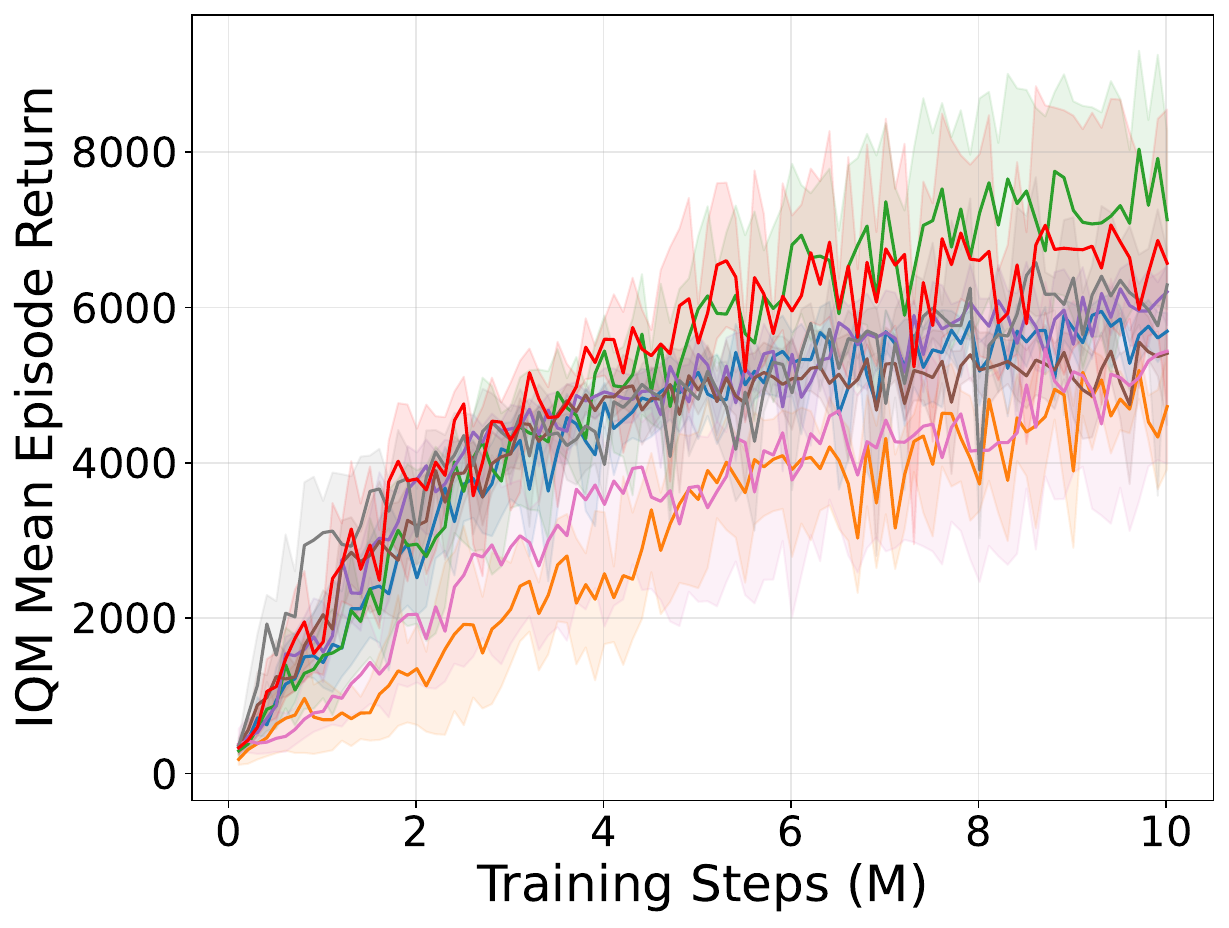}
        \captionsetup{skip=0pt}
        \caption{Walker2D-v2-2x3}
    \end{subfigure}%
    \hspace{0.5em}
    \begin{subfigure}[t]{0.25\textwidth}
        \centering
        \includegraphics[width=\textwidth]{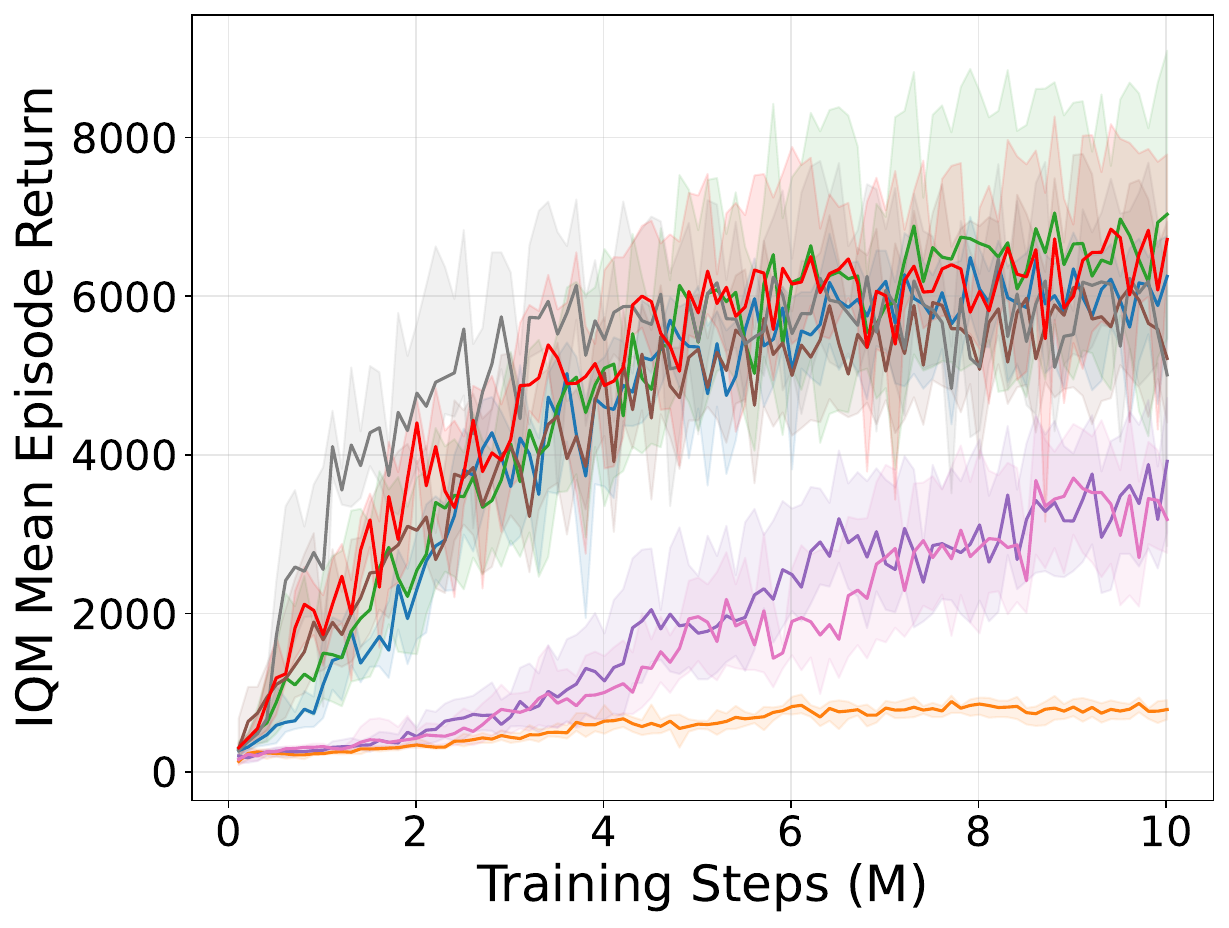}
        \captionsetup{skip=0pt}
        \caption{Walker2D-v2-6x1}
    \end{subfigure}%
    \hspace{0.5em}
    \begin{subfigure}[t]{0.25\textwidth}
        \centering
        \includegraphics[width=\textwidth]{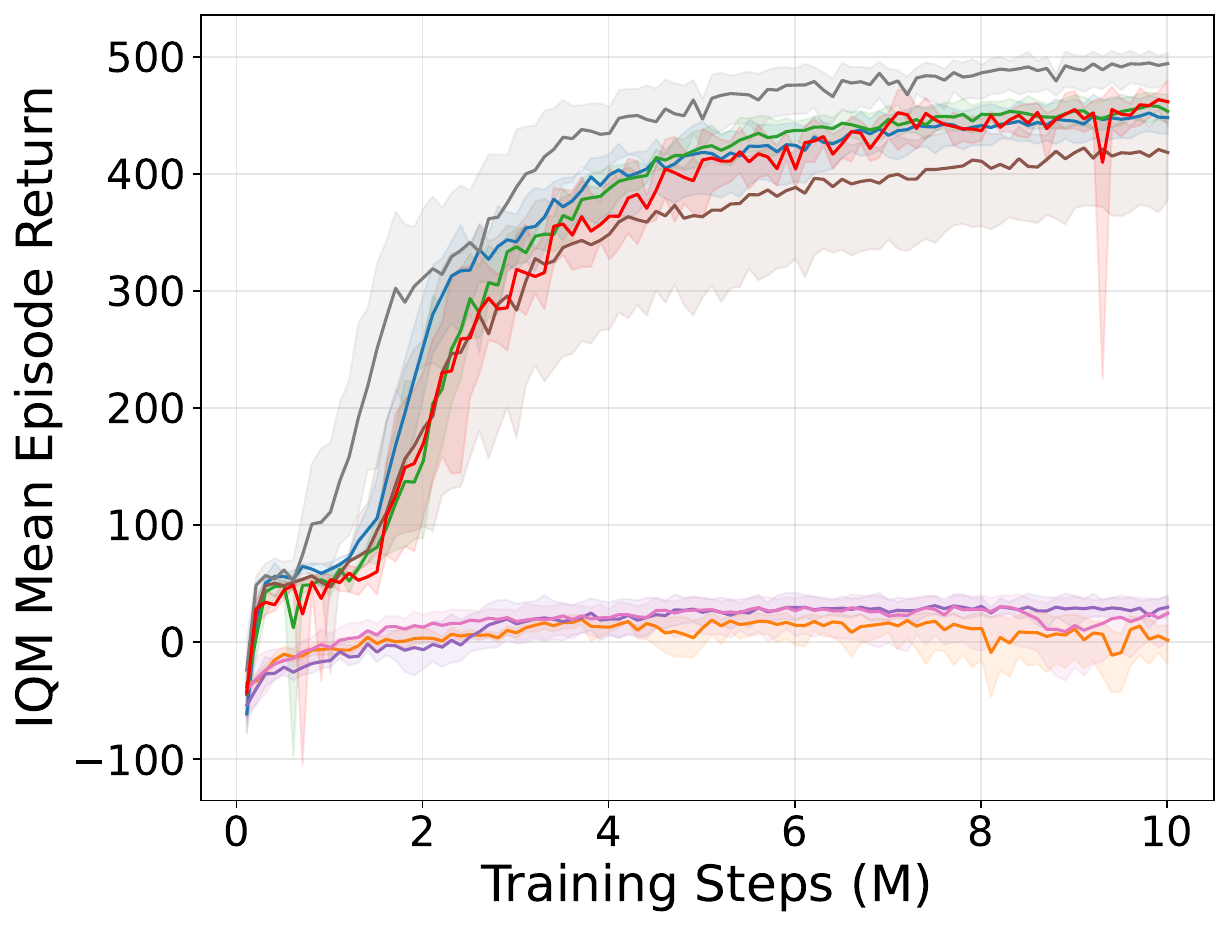}
        \captionsetup{skip=0pt}
        \caption{Swimmer-v2-10x2}
    \end{subfigure}
    
    % 전체 figure 캡션
    \captionsetup{skip=5pt}
    \caption{Performance evaluation on heterogeneous environment (MaMuJoCo).}
    \label{fig:mamujoco_performance}
\end{figure*}

% \begin{algorithm}[h!]
% \caption{Training Procedure of \textbf{Prism}}
% \label{alg:prism}
% \begin{small}
% \begin{algorithmic}[1]
% \Require Number of agents $N$, learning rate $\alpha$, replay buffer $\mathcal{D}$

% \State Initialize shared SVD-parameterized weights $(\mathbf{U}, \mathbf{s}, \mathbf{V})$
% \State Initialize agent-specific thresholds $\{\mathbf{t}_\Psi^{(i)}\}_{i=1}^N$
% \While{not converged}
%     \State Sample batch $(o, a, r, s')$ from $\mathcal{D}$
%     \For{each agent $i = 1, \ldots, N$}
%         \State Compute mask $\mathbf{m}_i = \mathrm{ReLU}\!\left( \mathbf{s}_{\text{norm}} - \sigma(\mathbf{t}_\Psi^{(i)}) \right)$
%         \State Compute $\mathbf{s}_i = \mathrm{concat}(\mathbf{s}_{common}, (\mathbf{s}_{separate} \odot \mathbf{m}_i))$
%         \State Construct policy parameters $\mathbf{W}_i = \mathbf{U} \, \mathrm{diag}(\mathbf{s}_i) \, \mathbf{V^\top}$
%         \State Compute policy loss $\mathcal{L}_i^{\text{policy}}$
%     \EndFor
%     \State Compute regularization losses $\mathcal{L}_{\text{div}}$ and $\mathcal{L}_{\text{ortho}}$
%     \State Update parameters:
%     \Statex \hspace{1em}$\theta \leftarrow \theta - \alpha \nabla_\theta \Big( \sum_i \mathcal{L}_i^{\text{policy}} + \lambda_{\text{div}}\mathcal{L}_{\text{div}} + \lambda_{\text{ortho}}\mathcal{L}_{\text{ortho}} \Big)$
% \EndWhile
% \end{algorithmic}
% \end{small}
% \end{algorithm}

\section{Experiments}
In this section, we empirically evaluate the effectiveness of the proposed method in multi-agent reinforcement learning.
First, we compare the proposed method with existing baselines to evaluate whether inducing diversity in the spectral space leads to consistent performance improvements (Sec. \ref{sec:exp_performance}).
Second, we evaluate the resource efficiency of \textbf{Prism} along two dimensions: 
performance robustness under limited parameter budgets and the growth of resource consumption as the number of agents increases (Sec.~\ref{sec:exp_resource_efficiency}).
Third, we conduct ablation studies to investigate the contributions of individual components in \textbf{Prism} to overall performance (Sec.~\ref{sec:exp_ablation}).
Finally, we visualize diversity in the spectral space, showing the emergence and differentiation of agent-specific spectral space during training (Sec.~\ref{sec:exp_visualization}).

%%%%%%%%%%%%%%%%%%%%%%%%%%%%%%%%%%%%%%%%%%%%%%%%%%%%%%%%%%%%%%%%%
\subsection{Experimental Setups}
\label{sec:experimental_setups}
We compare the proposed \textbf{Prism} against a wide range of parameter sharing strategies, covering approaches from no sharing to adaptive and structured methods:
\begin{itemize}[leftmargin=1.2em]
    \item \textbf{NoPS}: Independent networks for each agent, serving as a strong but parameter-inefficient baseline.  
    \item \textbf{FuPS} / \textbf{FuPS+ID}: All agents share the same network parameters. In FuPS+ID, each agent's input is augmented with one-hot agent identifiers for limited specialization.
    \item \textbf{SePS} \cite{pmlr-v139-christianos21a}: Cluster-level parameter sharing based on latent representations learned via a variational autoencoder, where agents within the same cluster share an identical network.
    \item \textbf{SNP} \cite{kim2023parameter}: Structured node-level pruning of a common dense network to form agent-specific subnetworks, achieving diversity through sparsity.
    \item \textbf{AdaPS} \cite{li2024adaptive}: Combines cluster-level grouping with structured pruning to leverage inter-agent similarity while enhancing intra-cluster diversity.
    \item \textbf{Kaleidoscope} \cite{NEURIPS2024_274d0146}: Applies learnable edge-level masks directly in weight space, enabling agents to discover distinct spaces end-to-end.
\end{itemize}

\begin{figure*}[t]
\centering

% ===== Legend =====
\includegraphics[width=0.5\textwidth]{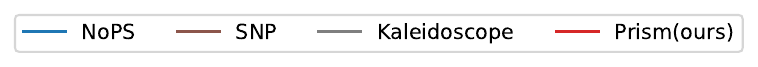}
% \vspace{0.8em}

% ===================== Row 1 : Full =====================
\begin{minipage}[t]{0.05\textwidth}
    \centering
    \raisebox{1.9\height}{\rotatebox{90}{\textbf{Full}}}
\end{minipage}%
\hfill
\begin{minipage}[t]{0.93\textwidth}
    \begin{subfigure}[t]{0.22\textwidth}
        \centering
        \includegraphics[width=\textwidth]{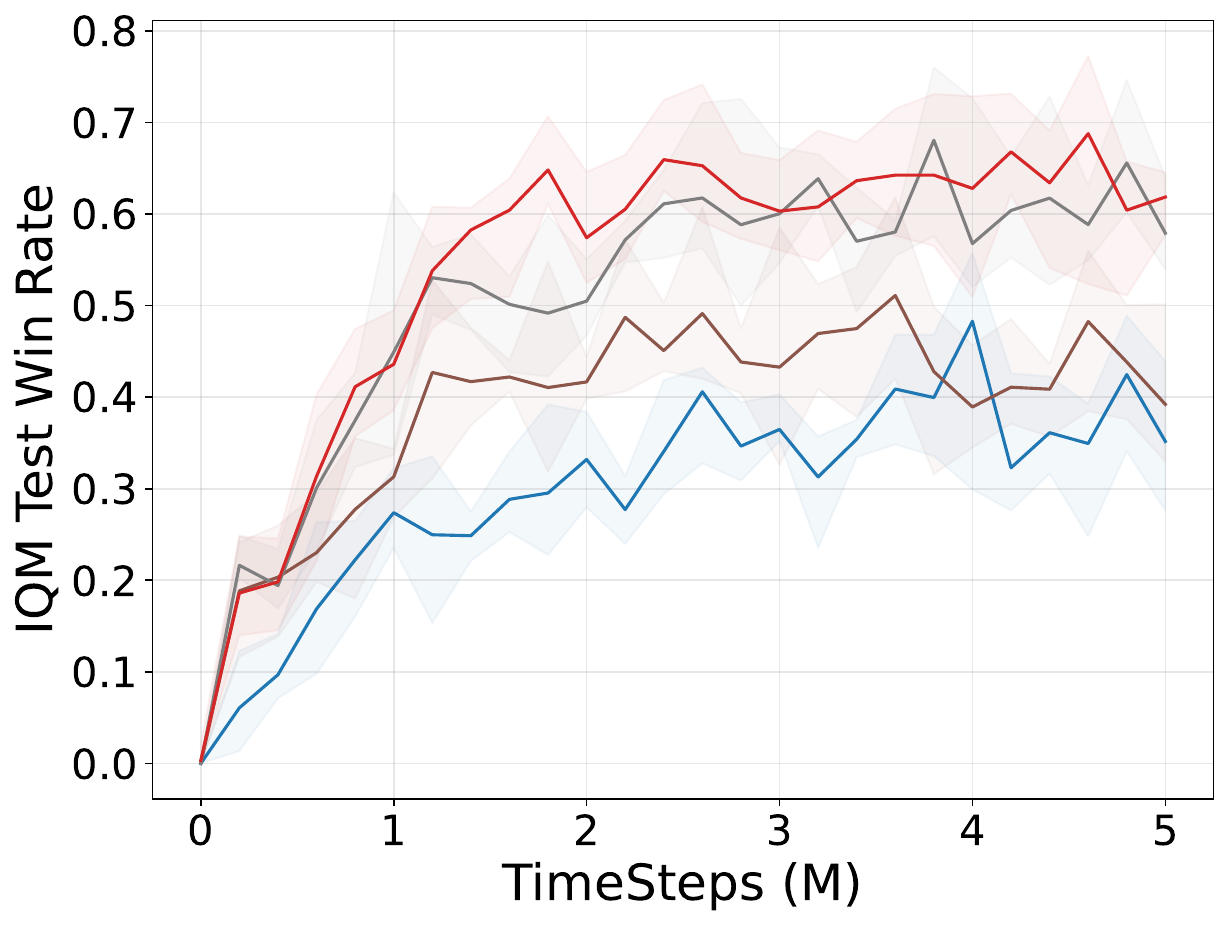}
    \end{subfigure}%
    \begin{subfigure}[t]{0.22\textwidth}
        \centering
        \includegraphics[width=\textwidth]{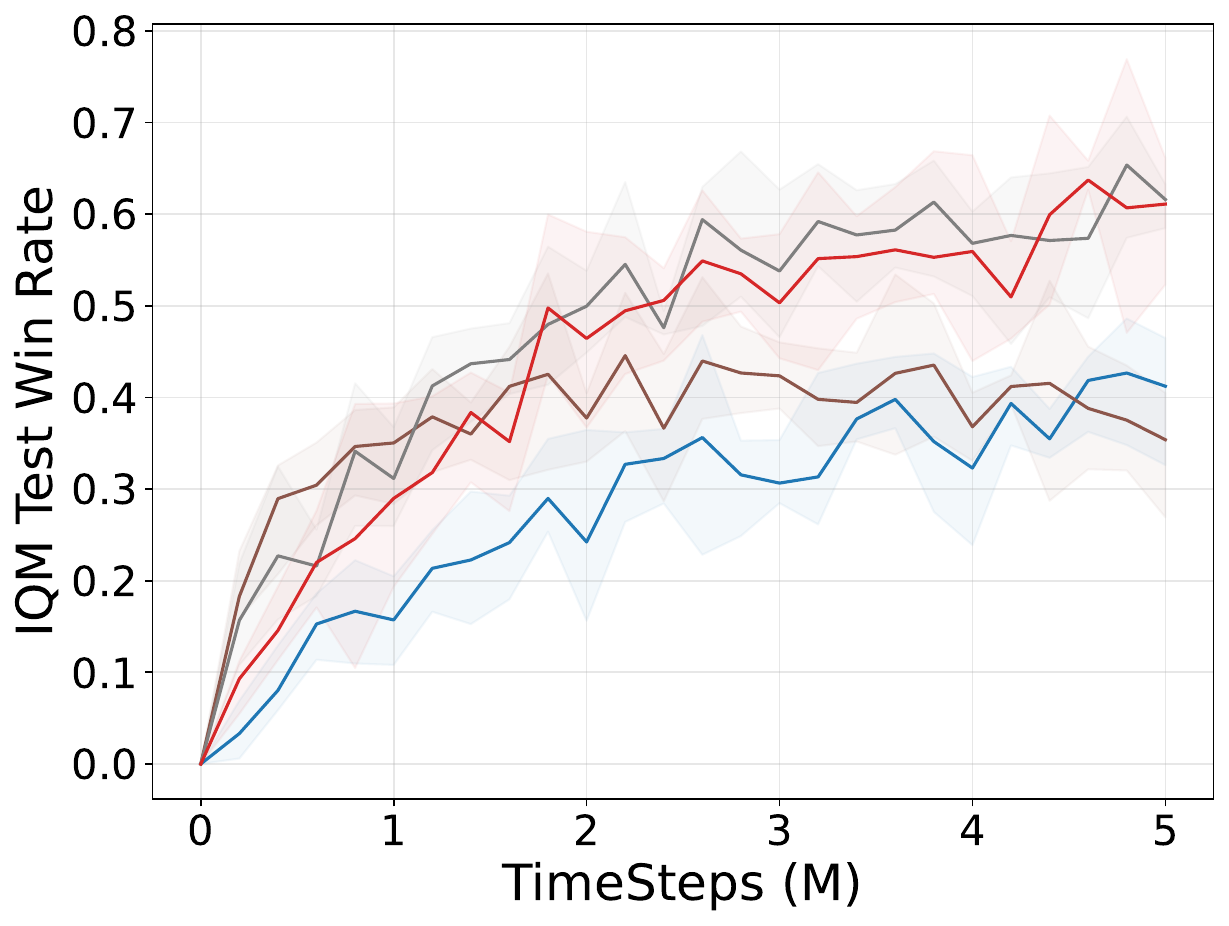}
    \end{subfigure}%
    \begin{subfigure}[t]{0.22\textwidth}
        \centering
        \includegraphics[width=\textwidth]{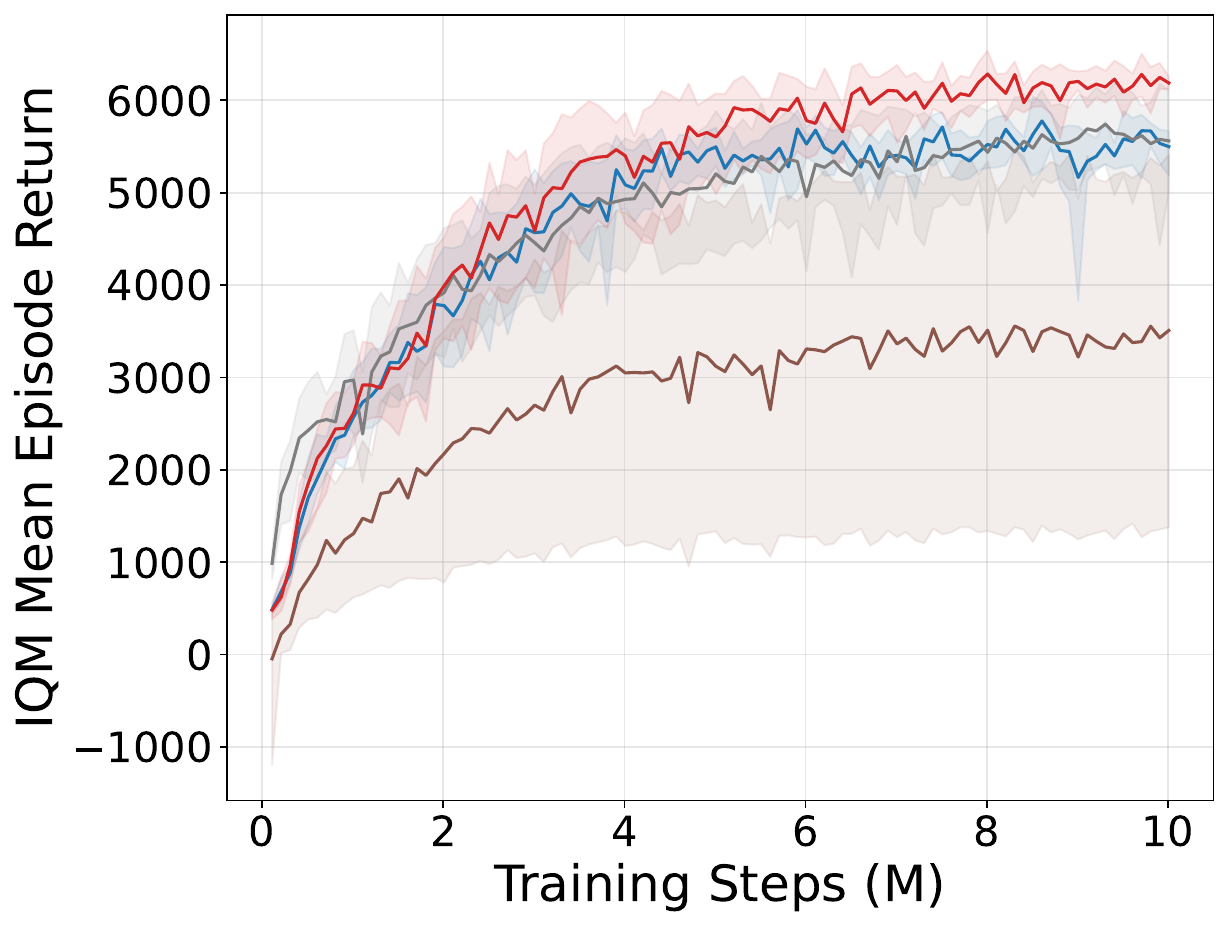}
    \end{subfigure}%
    \begin{subfigure}[t]{0.22\textwidth}
        \centering
        \includegraphics[width=\textwidth]{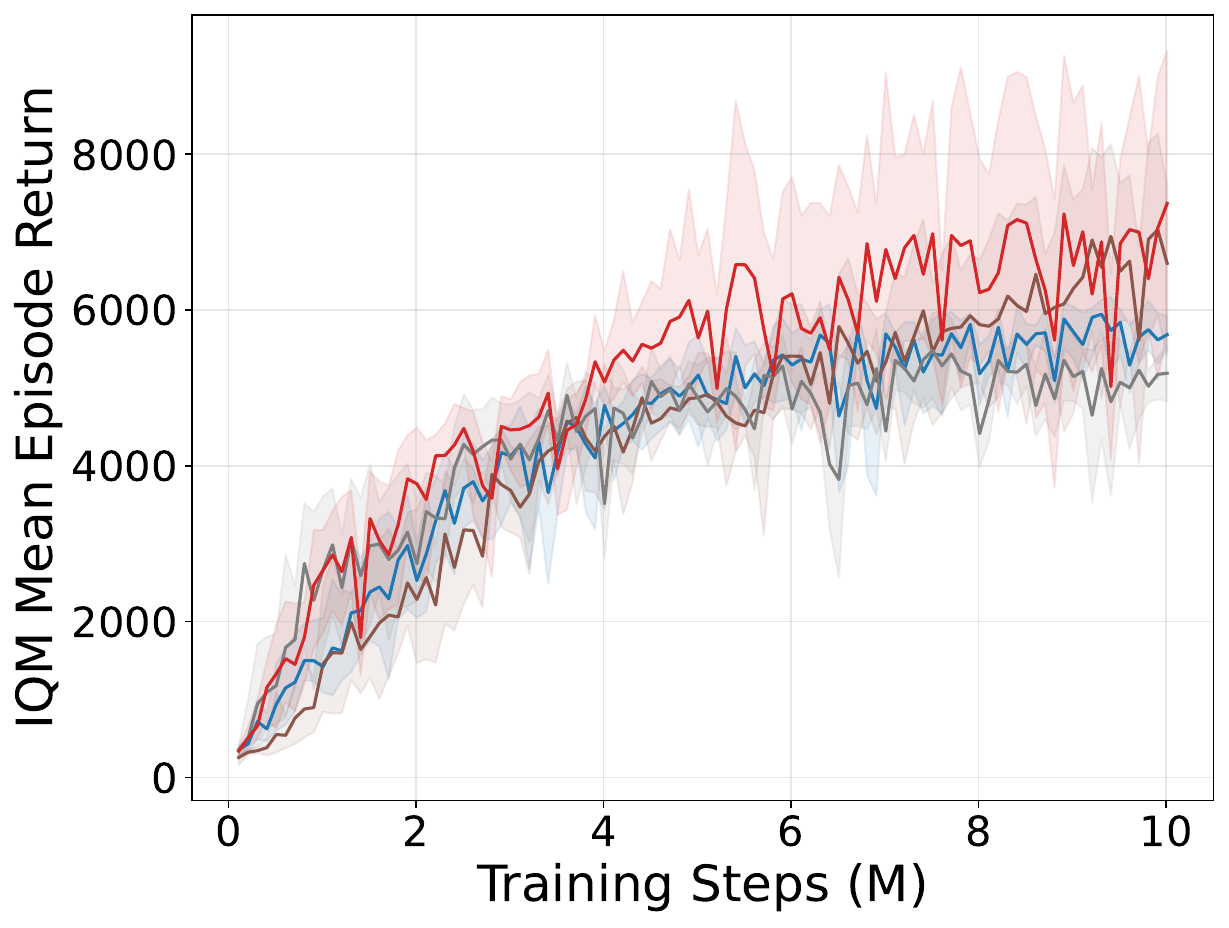}
    \end{subfigure}
\end{minipage}

% \vspace{1.2em}

% ===================== Row 2 : Half =====================
\begin{minipage}[t]{0.05\textwidth}
    \centering
    \raisebox{1.8\height}{\rotatebox{90}{\textbf{Half}}}
\end{minipage}%
\hfill
\begin{minipage}[t]{0.93\textwidth}
    \begin{subfigure}[t]{0.22\textwidth}
        \centering
        \includegraphics[width=\textwidth]{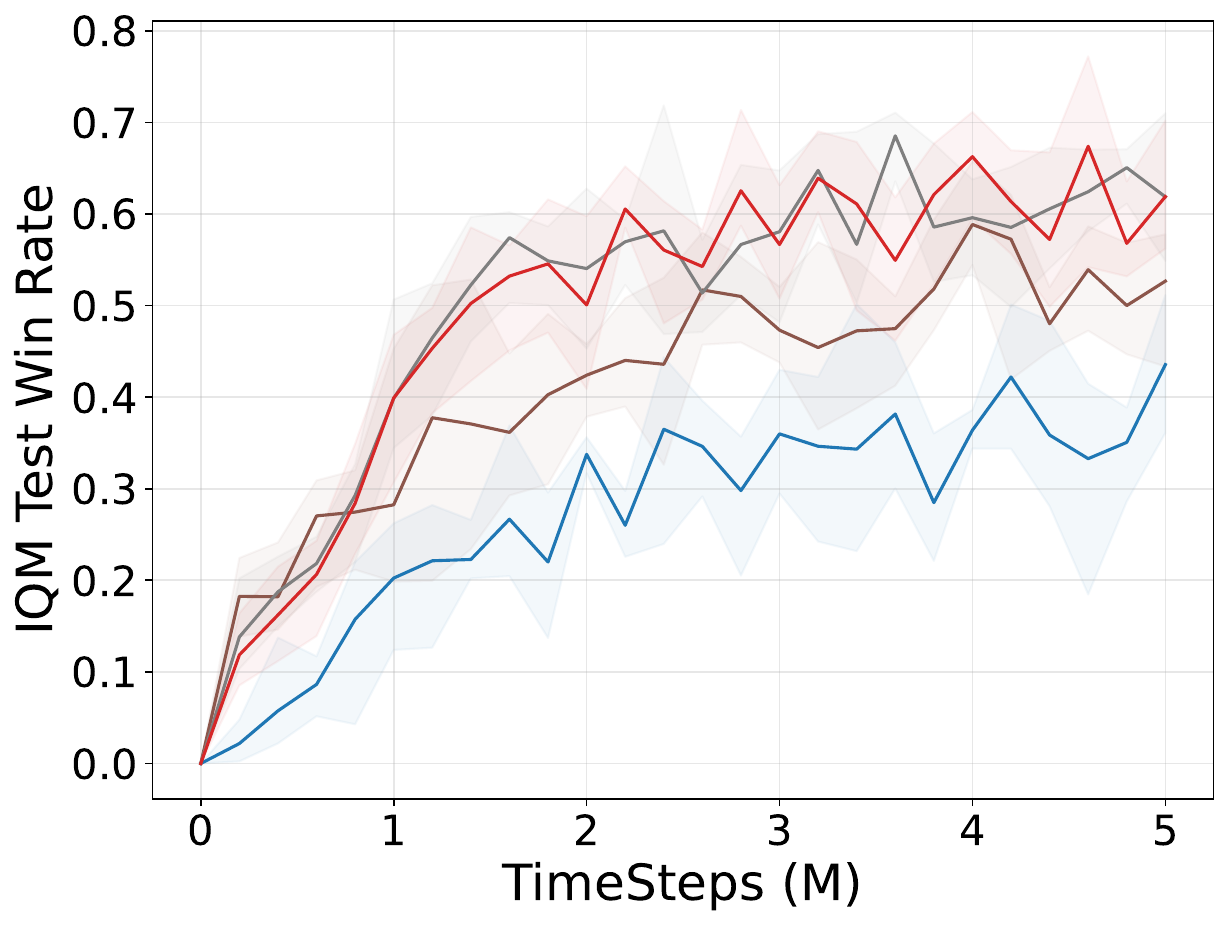}
        \caption{Terran 5\_vs\_5}
    \end{subfigure}%
    \begin{subfigure}[t]{0.22\textwidth}
        \centering
        \includegraphics[width=\textwidth]{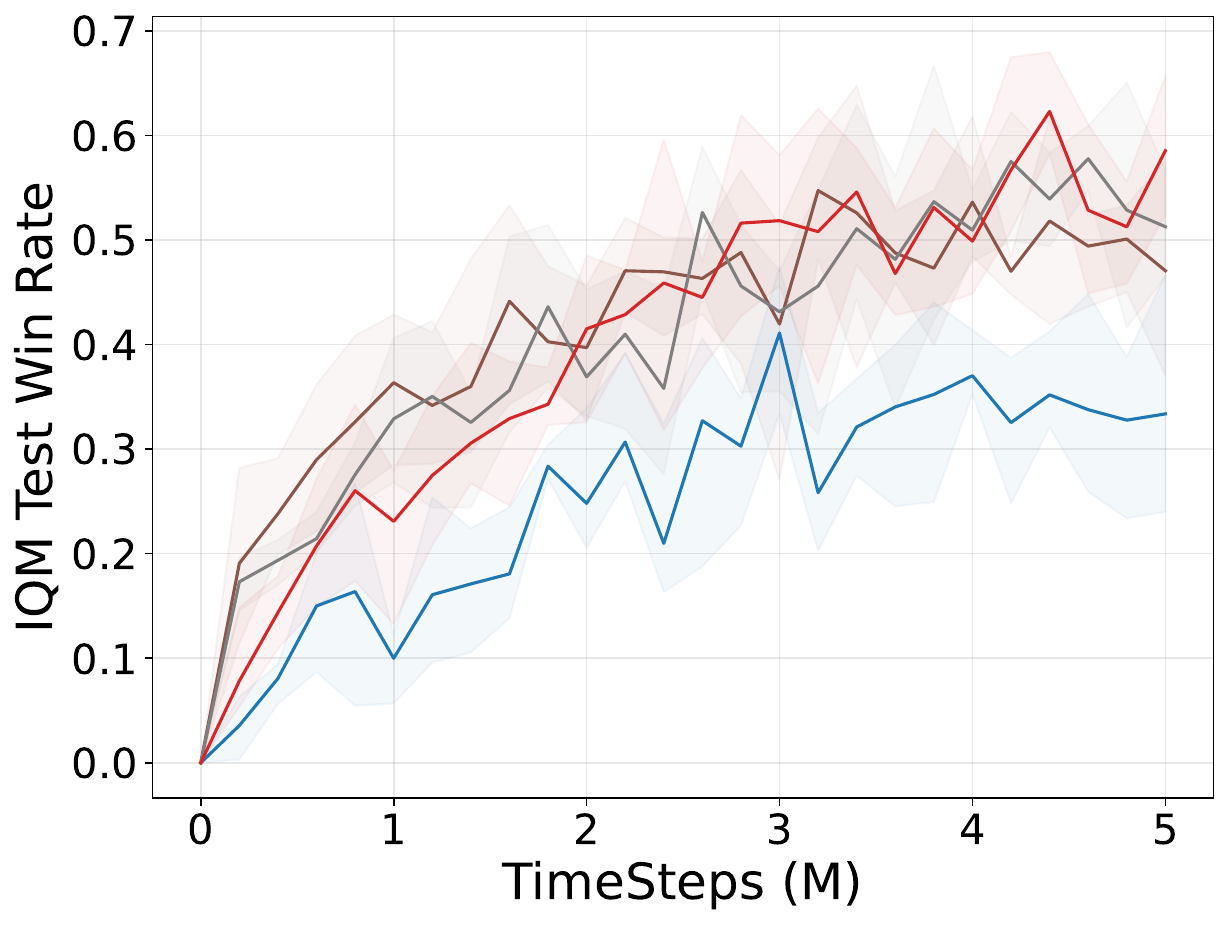}
        \caption{Protoss 5\_vs\_5}
    \end{subfigure}%
    \begin{subfigure}[t]{0.22\textwidth}
        \centering
        \includegraphics[width=\textwidth]{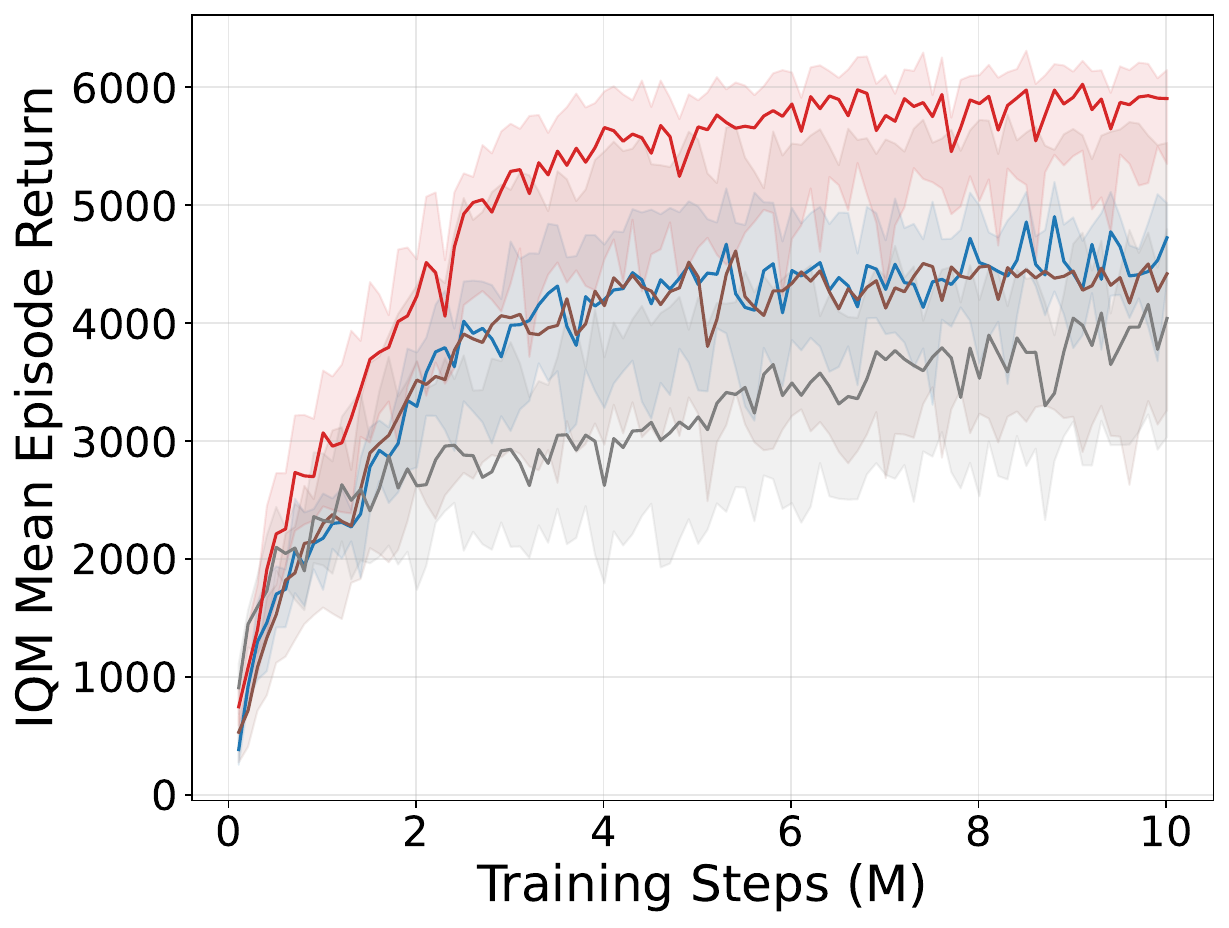}
        \caption{Ant-v2-4x2}
    \end{subfigure}%
    \begin{subfigure}[t]{0.22\textwidth}
        \centering
        \includegraphics[width=\textwidth]{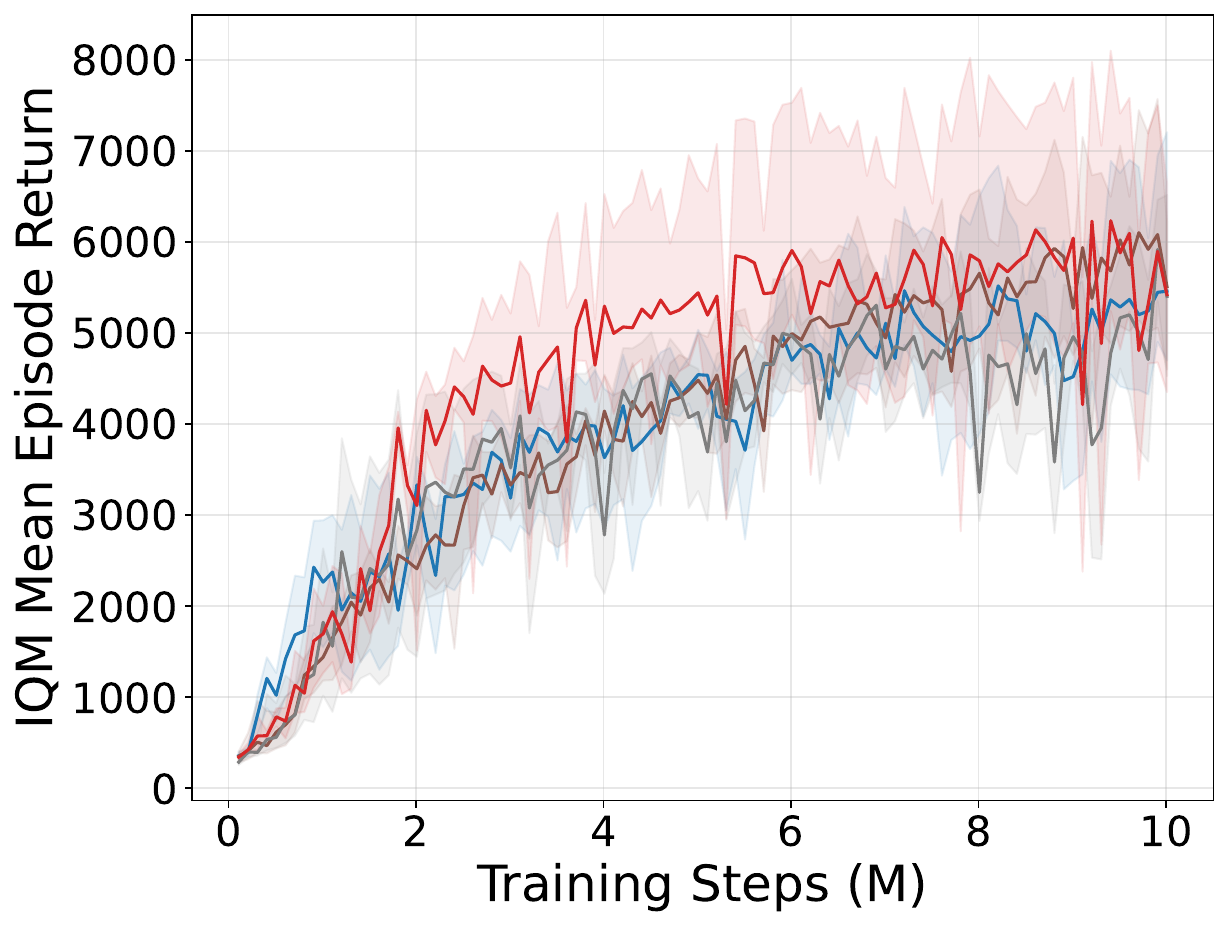}
        \caption{Walker2D-v2-2x3}
    \end{subfigure}
\end{minipage}

% \caption{Performance under parameter budget constraints on SMACv2 and MaMuJoCo.}
\caption{\textbf{Performance under parameter budget constraints on SMACv2 and MaMuJoCo}. Methods are evaluated with \textbf{Full} (matching NoPS parameters) and \textbf{Half} (50\% of NoPS) budgets by adjusting model width for a fair comparison.}
\label{fig:exp_efficiency}
\end{figure*}

We evaluate the performance of \textbf{Prism} across both homogeneous and heterogeneous multi-agent environments.
Specifically, we use three standard benchmarks: \textbf{Level-based Foraging (LBF)}~\cite{f82f11094573421ba6da695b9d6940c7}, \textbf{SMACv2}~\cite{ellis2023smacv} for discrete control under partial observation, and \textbf{MaMuJoCo}~\cite{peng2021facmac} for continuous control under full observation.
We adopt \textbf{QMIX}~\cite{pmlr-v80-rashid18a} for LBF and SMACv2 and \textbf{MATD3}~\cite{ackermann1910reducing} for MaMuJoCo, following the network configurations of \cite{NEURIPS2024_274d0146}.
All results are reported using the Interquartile Mean (IQM) of the mean episode return with 95\% bootstrap confidence intervals over five random seeds.

We use 3 clusters with unit-type masks in SMACv2 and 2 clusters with agent-level masks in LBF and MaMuJoCo, both with a pruning ratio of 0.1.
For \textbf{Prism}, we set the common ratio, orthogonal and diversity regularization coefficient to (0.6, 0.01, 5.0), (0.5, 0.01, 5.0), and (0.5, 0.01, 0.1) for LBF, SMACv2, and MaMuJoCo, respectively.
%%%%%%%%%%%%%%%%%%%%%%%%%%%%%%%%%%%%%%%%%%%%%%%%%%%%%%%%%%%%%%%
\subsection{Performance Evaluation}
\label{sec:exp_performance}
Before examining resource efficiency, we first evaluate the performance of \textbf{Prism} under standard training conditions. 
This provides a reference point for understanding how different parameter sharing strategies behave when network capacity is not explicitly considered. 
The performance curves are presented in Figure \ref{fig:smacv2_performance} (LBF, SMACv2) and Figure \ref{fig:mamujoco_performance} (MaMuJoCo).

\paragraph{\textbf{Homogeneous Setting.}}
In LBF and SMACv2, the optimal team strategy is largely homogeneous, as agents typically follow similar behaviors such as collectively targeting nearby objectives or focus-firing on the nearest enemy. As a result, learning efficiency of a shared policy becomes a critical factor under stochastic and partially observable dynamics.
As shown in Figure \ref{fig:smacv2_performance}, FuPS demonstrates higher performance than NoPS due to its superior sample efficiency under shared parameters. 
Clustering-based approaches such as SePS underperform in this setting due to their reliance on initial data, which limits adaptability to stochastic environments. 
Although AdaPS is also a clustering-based method, its use of a shared adaptive network improves sample efficiency, thereby leading to gradual performance improvements. 
In contrast, \textbf{Prism} achieves comparable or superior performance across maps, benefiting from a single shared network with learnable spectral masks that enable efficient adaptation to diverse agent behaviors.

\paragraph{\textbf{Heterogeneous Setting.}}
Unlike LBF and SMACv2, MaMuJoCo decomposes a single robot into multiple agents, where each agent controls distinct body parts with heterogeneous objectives (e.g., front and hind legs in \textit{HalfCheetah-v2-2x3}), given the same global observation.
Thus, inter-agent diversity is essential for achieving coordinated control. 
As shown in Figure \ref{fig:mamujoco_performance}, NoPS outperforms FuPS, highlighting the importance of diversity over strict parameter sharing in such heterogeneous tasks. 
Clustering-based methods (e.g., SePS, AdaPS) show inconsistent performance depending on the environment, as the number of optimal clusters varies across tasks—e.g., AdaPS performs well on \textit{HalfCheetah-v2-2x3} but underperforms elsewhere. 
\textbf{Prism}, by contrast, achieves consistently competitive and often superior performance across tasks, even outperforming Kaleidoscope, which applies learnable masks to critic networks as well.

%%%%%%%%%%%%%%%%%%%%%%%%%%%%%%%%%%%%%%%%%%%%%%%%%%%%%%%%%%%%%%%%%
\subsection{Resource Efficiency Evaluation}
\label{sec:exp_resource_efficiency}
\paragraph{\textbf{Under Resource Constraint.}}
In most real-world multi-agent systems, resource constraints such as limited computation or memory are unavoidable. 
Assessing parameter sharing in these conditions enables a more realistic evaluation of the trade-offs between memory efficiency and performance. 
Accordingly, we investigate different methods under progressively reduced parameter budgets, examining how effectively they exploit shared representations. 
We evaluate our method on SMACv2, using \textit{Terran 5\_vs\_5} and \textit{Protoss 5\_vs\_5} scenarios, and on MaMuJoCo with \textit{Ant-v2-4x2} and \textit{Walker2D-v2-2x3} tasks.

To simulate resource constraints, we define the parameter budget relative to NoPS.
We consider two settings: \textbf{Full}, where the parameter budget matches that of NoPS, and \textbf{Half}, where the budget is set to half of the NoPS parameter count.
For each method, the model width is adjusted to satisfy the specified budget, ensuring a fair comparison in terms of parameter count.

As shown in Figure \ref{fig:exp_efficiency}, the proposed method consistently maintains stable and competitive performance across different parameter budgets.
These results indicate that spectral-space masking enables more effective parameter sharing and provides a clear advantage in resource-limited multi-agent systems.

\begin{figure}[t]
    \centering
    % 상단 그래프 이미지
    \includegraphics[width=1.0\linewidth]{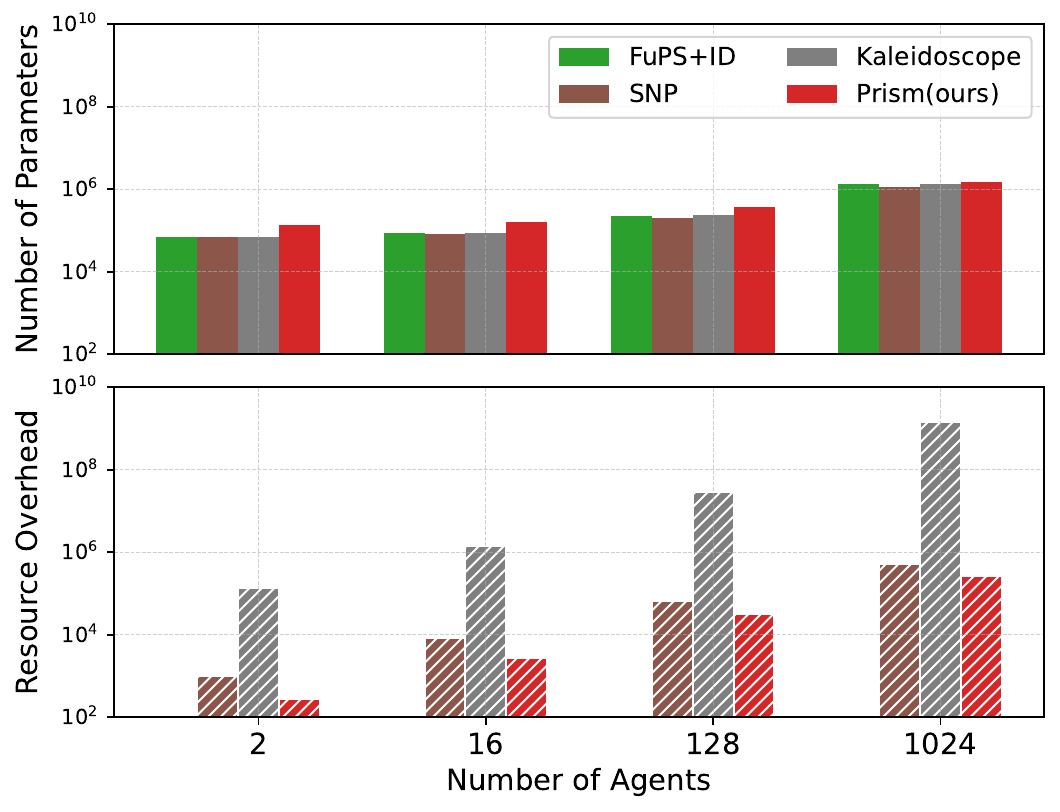}

    % \vspace{0.5em} % 이미지와 표 사이 간격

    % 표 시작 (resizebox로 가로폭 맞춤)
    % \resizebox{\linewidth}{!}{%
    \resizebox{0.9\linewidth}{!}{%
        \begin{tabular}{ccccc} % 수직선(|) 제거
            \toprule % 맨 위 굵은 선
            \textbf{Agents} & \textbf{FuPS+ID} & \textbf{SNP} & \textbf{Kaleidoscope} & \textbf{Prism} \\
            \midrule % 헤더 아래 중간 선
            2    & 0.0000 & 0.0145 & 0.6618 & \textbf{0.0020} \\
            16   & 0.0000 & 0.0892 & 0.9402 & \textbf{0.0166} \\
            128  & 0.0000 & 0.2488 & 0.9922 & \textbf{0.0829} \\
            1024 & 0.0000 & 0.3204 & 0.9990 & \textbf{0.1483} \\
            \bottomrule % 맨 아래 굵은 선
        \end{tabular}%
    }

    \caption{
    Resource efficiency evaluation with respect to the number of agents.
    \textbf{Top:} The total number of parameters and the additional resource overhead required by each method as the number of agents increases.
    \textbf{Bottom:} The normalized resource overhead, defined as the ratio of additional resources to the total model size (resource / (parameters + resource)).
    }
    \label{fig:agent_scaling}
\end{figure}

\paragraph{\textbf{Resource Efficiency with Agent Scaling}}
To further investigate the scalability of parameter sharing frameworks, we analyze how memory resource changes with the number of agents using MaMuJoCo (\textit{Manyagent Swimmer}). As shown in Figure \ref{fig:agent_scaling}, the total number of parameters grows only marginally with the number of agents across all methods, since all approaches rely on parameter sharing. In contrast, the resource overhead exhibits a substantially higher growth rate, as per-agent components such as node-level or edge-level masks are instantiated independently for each agent. Therefore, reducing resource overhead is essential to scalability in multi-agent systems.

Kaleidoscope leads to a significant resource overhead because each agent maintains a full edge mask corresponding to the weight space, resulting in resource growth that far exceeds the increase in shared parameters. SNP, which assigns node-level masks per agent, reduces this overhead compared to Kaleidoscope but still suffers from a rapidly increasing resource usage as the number of agents scales.

By contrast, \textbf{Prism} achieves substantially improved resource efficiency by leveraging a more compact spectral space, where spectral masks replace node- or edge-level agent-specific masks in the original weight space. This design significantly mitigates the growth of resource overhead while preserving scalability, making \textbf{Prism} more suitable for large-scale multi-agent systems. Quantitatively, Prism consistently exhibits a lower normalized resource overhead than SNP and Kaleidoscope (Figure \ref{fig:agent_scaling}, bottom).

%%%%%%%%%%%%%%%%%%%%%%%%%%%%%%%%%%%%%%%%%%%%%%%%%%%%%%%%%%%%%%%%%
\subsection{Ablation Study}
\label{sec:exp_ablation}
We perform ablation studies to investigate the contribution of individual design choices in two representative environments: SMACv2 (\textit{Terran\_5\_vs\_5}) and MaMuJoCo (\textit{Ant-v2-4x2}). Our analysis focuses on the robustness of the proposed regularization terms and the effectiveness of diversity.

\begin{figure}[h!]
    \centering
    \vspace{0.5em}
    
    \begin{subfigure}[t]{0.23\textwidth}
        \centering
        \includegraphics[width=\textwidth]{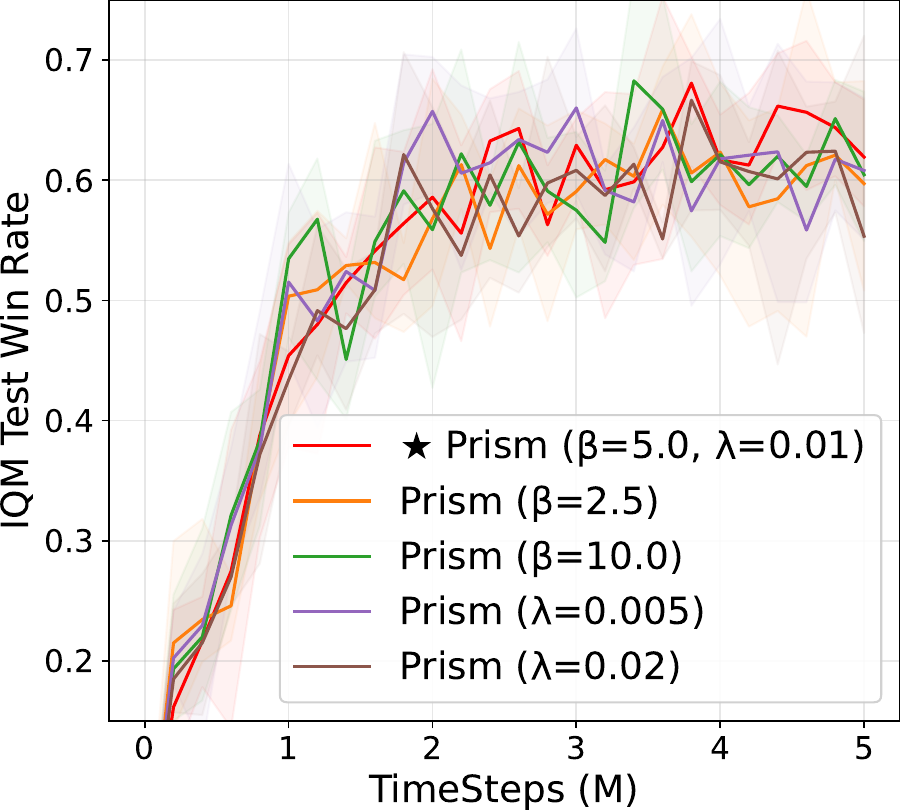}
        \captionsetup{skip=0pt}
        \caption{Terran\_5\_vs\_5}
        \label{fig:ablation_reg_smacv2}
    \end{subfigure}%
    \hspace{0.2em}
    \begin{subfigure}[t]{0.23\textwidth}
        \centering
        \includegraphics[width=\textwidth]{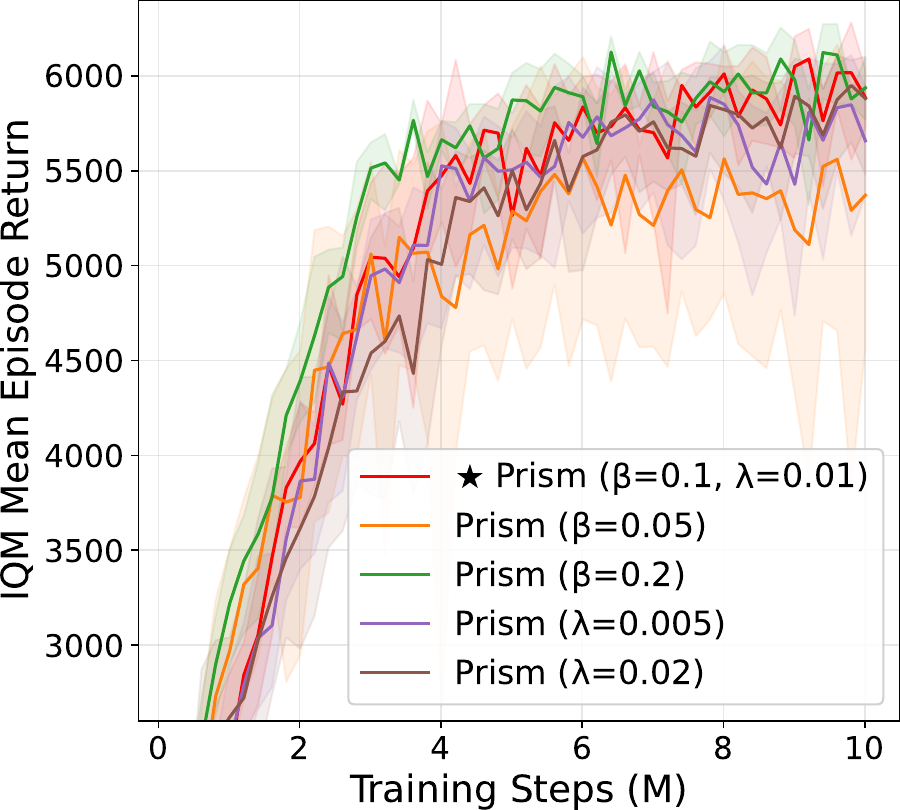}
        \captionsetup{skip=0pt}
        \caption{Ant-v2-4x2}
        \label{fig:ablation_reg_mamujoco}
    \end{subfigure}%
    \captionsetup{skip=1pt}
    \caption{Ablation study on the regularization strengths of diversity ($\beta$) and orthogonality ($\lambda$).}
    \label{fig:exp_ablation_reg}
\end{figure}

\paragraph{\textbf{Regularization Robustness.}}
We first examine the sensitivity of performance to the strength of orthogonal and diversity regularization. Specifically, we experiment with coefficients set to half and twice the default hyperparameter values used in Sec.~\ref{sec:exp_performance}. As shown in Figure \ref{fig:exp_ablation_reg}, the impact of regularization strength differs between diversity and orthogonal regularization.
In particular, the effect of diversity regularization is more pronounced in heterogeneous environments such as MaMuJoCo. Reducing the diversity coefficient to half of the default value leads to degraded performance, whereas doubling the coefficient results in improved final performance. This suggests that increasing diversity regularization can be beneficial in environments where agent heterogeneity plays a critical role.
In contrast, orthogonal regularization exhibits negligible performance differences across all environments and coefficient settings, indicating low sensitivity to its strength.
Overall, these results show that \textbf{Prism} is generally robust to regularization hyperparameters, while highlighting that stronger diversity regularization may offer additional benefits in heterogeneity-critical tasks.

\begin{figure}[h!]
    \centering
    \begin{subfigure}[t]{0.23\textwidth}
        \centering
        \includegraphics[width=\textwidth]{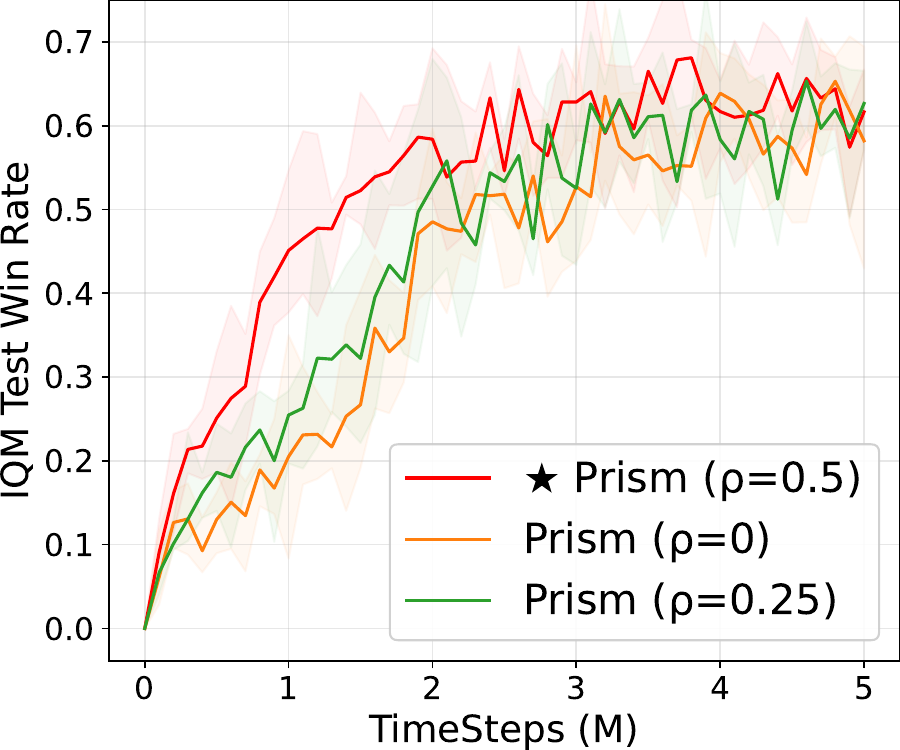}
        \captionsetup{skip=0pt}
        \caption{Terran\_5\_vs\_5}
        \label{fig:ablation_common_smacv2}
    \end{subfigure}%
    \hspace{0.2em}
    \begin{subfigure}[t]{0.23\textwidth}
        \centering
        \includegraphics[width=\textwidth]{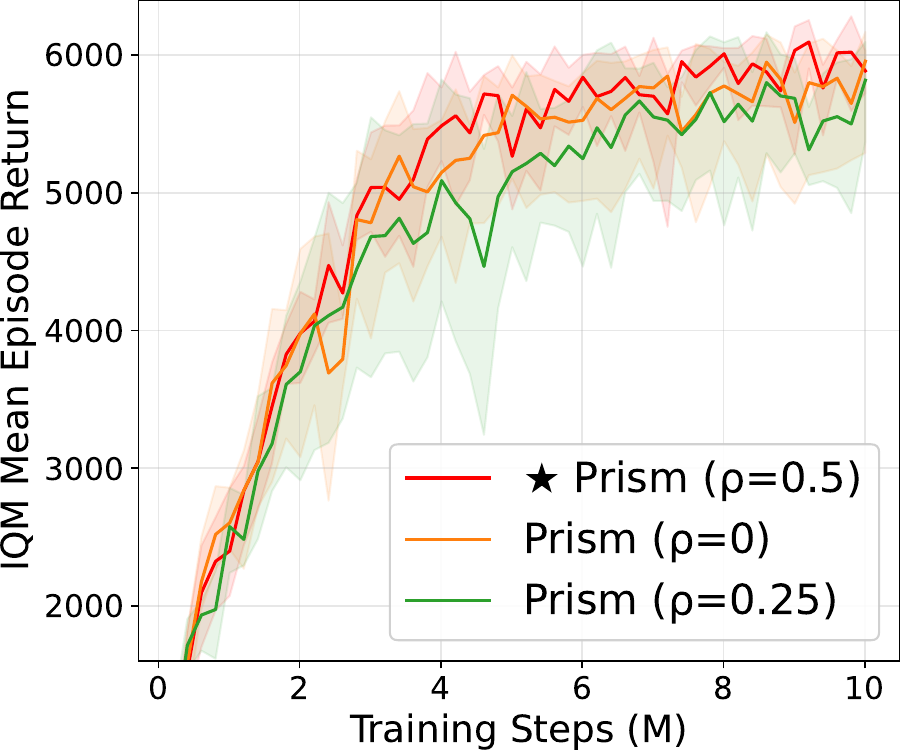}
        \captionsetup{skip=0pt}
        \caption{Ant-v2-4x2}
        \label{fig:ablation_common_mamujoco}
    \end{subfigure}%
    \captionsetup{skip=1pt}
    \caption{Ablation study on common mask ratio $\rho$.}
    \label{fig:exp_ablation_common}
\end{figure}

\paragraph{\textbf{Spectral Sharing Ratio.}}
We further analyze the sensitivity of \textbf{Prism} to the ratio between shared and agent-specific spectral components. To this end, we adjust the proportion of shared components among {0.0, 0.25, 0.5} and evaluate the final performance across representative environments.
As shown in Figure \ref{fig:exp_ablation_common}, in \textit{Terran\_5\_vs\_5}, a higher proportion of shared spectral components accelerates convergence. This trend aligns with the characteristics of the task, where sample efficiency is critical, and a higher degree of common ratio results in more stable gradient flow and improved exploitation of shared learning signals.
In contrast, \textit{Ant-v2-4x2} exhibits minimal sensitivity to the sharing ratio, indicating that in highly agent-specific environments, performance is primarily governed by agent-specific specialization rather than inter-agent spectral sharing.
Overall, these findings suggest that environments emphasizing sample efficiency benefit from a greater proportion of shared spectral components, while in heterogeneous coordination tasks, performance improvements are primarily achieved through stronger diversity regularization rather than adjusting the sharing ratio.

% \input{figures/ablation_mask}
% \paragraph{\textbf{Spectral Masking.}}
% To ensure that the performance gain is not merely due to SVD-based training itself, we additionally consider a variant, \textbf{FuPS+ID w/ SVD}, where both SVD training and orthogonal regularization are applied but spectral masking is disabled. As shown in Figure ~\ref{fig:exp_ablation_mask}, this variant shows moderate improvement but still underperforms Prism across environments. Notably, in \textbf{Ant-v2-4x2}, removing spectral masking increases variance, suggesting that spectral diversity enables inter-agent specialization, leading to more stable and improved performance.

%%%%%%%%%%%%%%%%%%%%%%%%%%%%%%%%%%%%%%%%%%%%%%%%%%%%%%%%%%%%%%%%%
\subsection{Visualization of Diversity in the Spectral Space}
\label{sec:exp_visualization}

\begin{figure*}[t!]
    \centering
    % \vspace{0.5em}
    
    \begin{subfigure}[t]{0.67\textwidth}
        \centering
        \includegraphics[width=\textwidth]{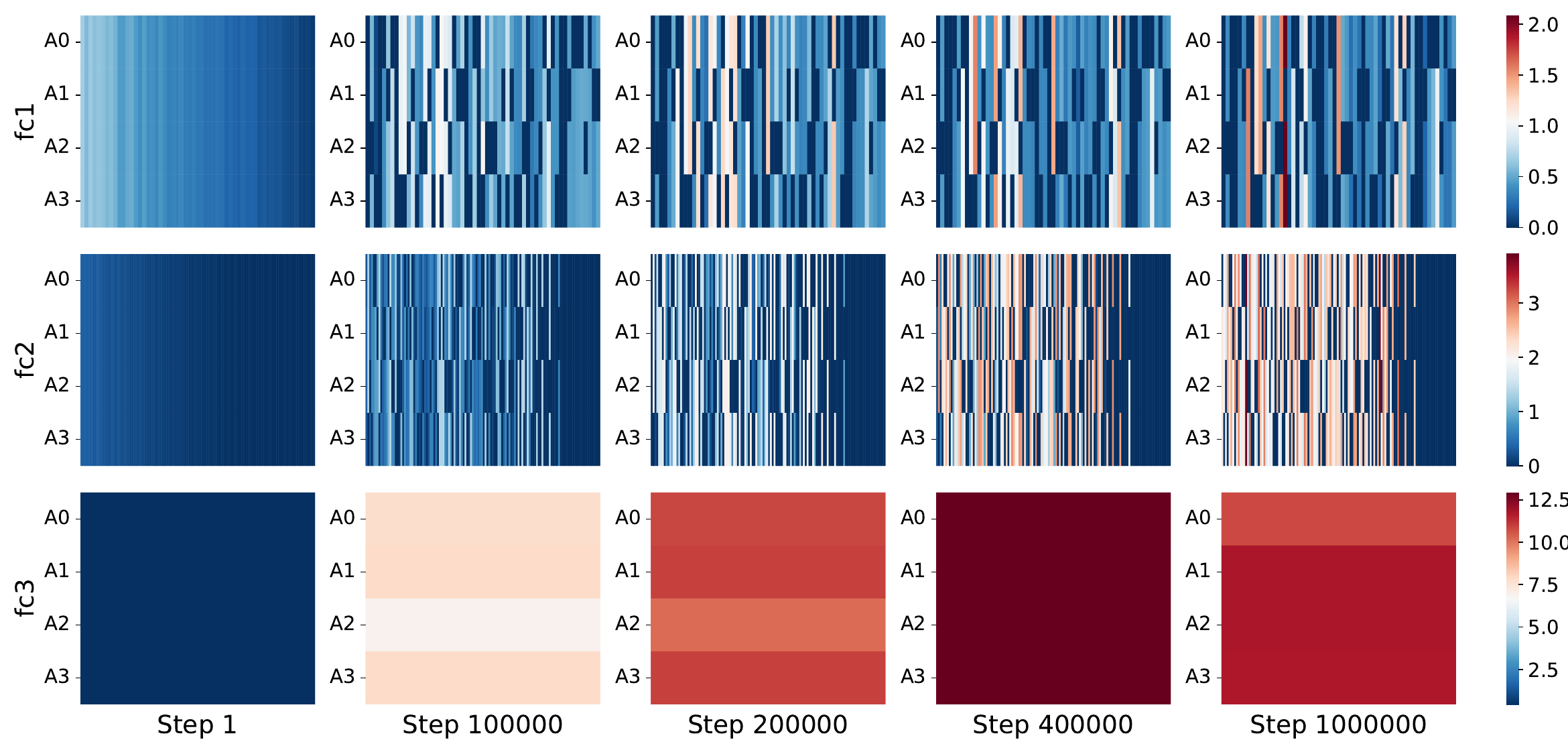}
        \captionsetup{skip=0pt}
        \caption{Agent-Specific Singular Value}
        \label{fig:analysis_spectral_diversity}
    \end{subfigure}%
    \hspace{0.2em}
    \begin{subfigure}[t]{0.32\textwidth}
        \centering
        \includegraphics[width=\textwidth]{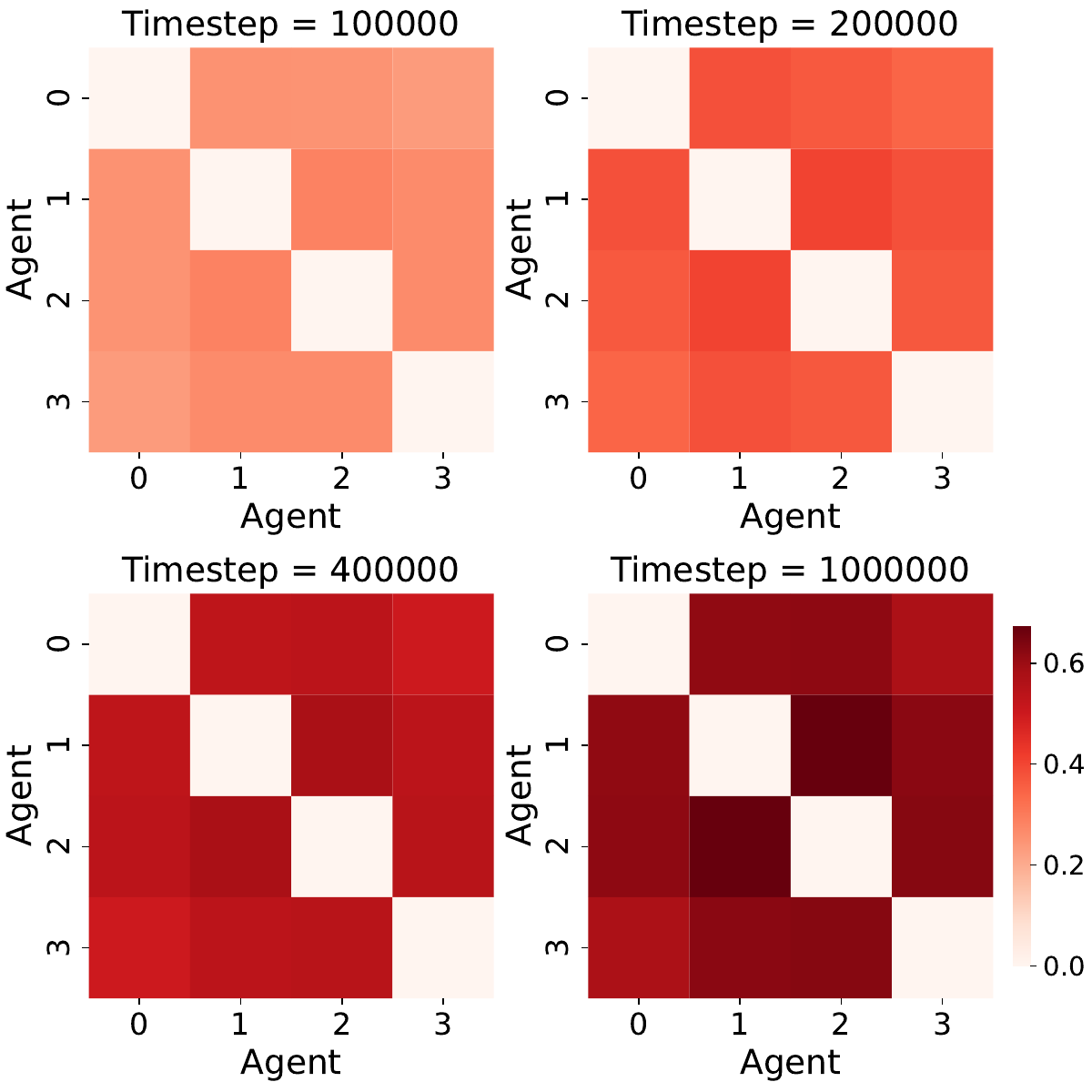}
        \captionsetup{skip=0pt}
        \caption{Pairwise Mask Differences}
        \label{fig:analysis_mask_differences}
    \end{subfigure}%
    \captionsetup{skip=1pt}
    % \caption{Visualization of emergent diversity across agents during training.}
    \caption{\textbf{Visualization of Agent Representations in the Spectral Space during Training (Ant-v2-4x2).} (a) Heatmaps of masked singular values (spectral masks) for each agent at different training steps. (b) Pairwise mask-difference matrices, showing increasing inter-agent specialization over time.}
    \label{fig:exp_analysis_masking}
\end{figure*}

We next examine how diversity emerges during training. 
Figure \ref{fig:exp_analysis_masking} (a) visualizes the evolution of masked singular values for each agent as a heatmap on \textit{Ant-v2-4$\times$2}. 
At the early stage of training, agents attend to similar spectral components, resulting in limited diversity. 
As learning progresses, the spectral masks gradually diverge, assigning distinct subsets of singular directions to different agents. 
This specialization indicates that agents adaptively occupy agent-specific spectral subspaces while still leveraging shared representational bases. 
Quantitatively, Figure~\ref{fig:exp_analysis_masking} (b) reports the pairwise L1 distances between agent-specific spectral masks over training.
The increasing magnitude of these mask differences indicates that agents progressively allocate distinct weights to shared singular components, resulting in increasingly differentiated spectral utilization across agents.
% Quantitatively, the pairwise L1 distance between agent-specific singular subspace increases over time, indicating that their spectral representations become increasingly distinct within the spectral space.
% These findings demonstrate that \textbf{Prism}'s learnable spectral masking preserves scalability while encouraging emergent inter-agent specialization, leading to efficient and diverse policy representations.

\section{Related Works}

\subsection{Parameter Sharing Methods}
Parameter sharing~\cite{tan1993multi} is widely used in MARL to improve scalability, but a single shared policy often struggles to represent heterogeneous agent behaviors. 
To address this, prior work introduces policy heterogeneity through clustering or structured sparsity. 
SePS~\cite{pmlr-v139-christianos21a} clusters agents based on latent representations, SNP-PS~\cite{kim2023parameter} induces diversity via structured pruning, and AdaPS~\cite{li2024adaptive} combines clustering with pruning for efficient sharing. 
More recently, Kaleidoscope~\cite{NEURIPS2024_274d0146} employs learnable weight-space masks to enable agent-specific subspaces within a shared network. 
Despite these advances, existing methods still face a trade-off between memory efficiency and heterogeneity.
In contrast, inducing diversity in the spectral space can offer a way to alleviate this trade-off with reduced parameter overhead.

\subsection{SVD Training in Deep Learning}
% SVD-based training has been explored in a few studies in deep learning to induce low-rank structure and improve efficiency.
% Yang et al.~\cite{Yang_2020_CVPR_Workshops} proposed to explicitly optimize neural networks in their singular value decomposition form, introducing orthogonality regularization on singular vectors and sparsity-inducing penalties on singular values to encourage low-rank representations without repeated SVD computations. 
% Building upon this direction, Khodak et al.~\cite{khodak2021initialization} studied initialization and regularization strategies for factorized neural layers, proposing spectral initialization and Frobenius decay to stabilize optimization and enhance generalization. 
% These approaches primarily focus on model compression and efficiency by reducing the effective rank of weight matrices. 

% In contrast, our work leverages SVD training not with the primary goal of sparsity or low-rank compression, but to induce spectral diversity in multi-agent reinforcement learning. By applying learnable spectral masking, we adaptively allocate agents to distinct spectral subspaces, thereby achieving inter-agent diversity while maintaining scalability. 

SVD-parameterizations control the effective rank of neural layers by operating in the spectral domain. ~\cite{Yang_2020_CVPR_Workshops} parameterize weights in SVD form and promote low rank via sparsity over singular values, enabling singular-value pruning. ~\cite{khodak2021initialization} show that SVD-informed initialization and regularization make low-rank factorized layers train reliably across architectures (including attention). In a complementary compression setting, ~\cite{wang2024unstructured} use truncated SVD to convert unstructuredly pruned sparse weights into dense low-rank factorisations, with regularization and learnable singular-value selection to adapt truncation per matrix.
In contrast, our work leverages SVD training not with the primary goal of sparsity or low-rank compression, but to induce agent differentiation in the spectral space for multi-agent reinforcement learning.

\section{Conclusion}

In this paper, we introduced \textbf{Prism}, a novel parameter sharing framework for multi-agent reinforcement learning that induces diversity in the spectral space through SVD-based factorization and learnable spectral masking. By representing shared weights in the spectral domain, Prism effectively decomposes the parameter space into shared and agent-specific spectral subspaces, achieving both memory efficiency and heterogeneity without redundant parameters. 
Extensive experiments across homogeneous (LBF, SMACv2) and heterogeneous (MaMuJoCo) benchmarks demonstrated that \textbf{Prism} consistently achieves competitive or superior performance compared to state-of-the-art parameter sharing approaches, particularly under resource-constrained settings where memory efficiency and heterogeneity are both critical. 
Ablation studies further show that varying the diversity regularization benefits heterogeneous environments, while increasing the common ratio is more effective in homogeneous settings.

% While \textbf{Prism} shows strong generality and scalability, some limitations remain. 
While \textbf{Prism} effectively addresses scalability challenges and demonstrates strong generality across diverse environments, some limitations remain.
First, the approach introduces additional hyperparameters for controlling the ratio between shared and agent-specific spectral components, which require task-dependent tuning. 
Second, our experiments were conducted exclusively in off-policy settings; extending \textbf{Prism} to on-policy algorithms (e.g., MAPPO) is an interesting future direction that may further enhance adaptability and stability in dynamic environments.

\appendix
\section{Training Procedure}
\begin{algorithm}[H]
\caption{Training Algorithm for \textbf{Prism}}
\label{alg:prism}
\textbf{Input:} Number of agents $N$, learning rate $\alpha$, replay buffer $\mathcal{D}$\\
\textbf{Output:} Updated policy $\theta$
\begin{algorithmic}[1]
\STATE Initialize shared SVD-parameterized weights $(\mathbf{U}, \mathbf{s}, \mathbf{V})$
\STATE Initialize agent-specific thresholds $\{\mathbf{t}_{\Psi}^{(i)}\}_{i=1}^{N}$
\WHILE{not converged}
    \STATE Sample a minibatch $(o,a,r,o')$ from $\mathcal{D}$
    \FOR{each agent $i = 1,\ldots,N$}
        \STATE Compute mask $\mathbf{m}_i \leftarrow \mathrm{ReLU}\!\left(
        \mathbf{s}_{\mathrm{norm}} - \sigma(\mathbf{t}_{\Psi}^{(i)}) \right)$
        \STATE Compute
        $\mathbf{s}_i \leftarrow \mathrm{concat}(
        \mathbf{s}_{\mathrm{common}},
        \mathbf{s}_{\mathrm{separate}} \odot \mathbf{m}_i )$
        \STATE Construct policy parameters
        $\mathbf{W}_i \leftarrow \mathbf{U}\,\mathrm{diag}(\mathbf{s}_i)\,\mathbf{V}^{\top}$
        \STATE Compute policy loss $\mathcal{L}_i^{\mathrm{policy}}$
    \ENDFOR
    \STATE Compute diversity regularization
    $\mathcal{J}_{\mathrm{div}}$
    \STATE Compute orthogonality loss
    $\mathcal{L}_{\mathrm{ortho}}$
    \STATE 
    $\theta \leftarrow \theta - \alpha \nabla_{\theta}
    \Big(\sum_i \mathcal{L}_i^{\mathrm{policy}}
    - \lambda_{\mathrm{div}}\mathcal{J}_{\mathrm{div}}
    + \lambda_{\mathrm{ortho}}\mathcal{L}_{\mathrm{ortho}}\Big)$
\ENDWHILE
\end{algorithmic}
\end{algorithm}

\section{Comparison of Parameter Efficiency: Weight vs Spectral Masking}
To provide a closed-form comparison of memory usage between weight-space masking and spectral-space masking, we analyze the common case where hidden-layer weight matrices are square of size $d \times d$. This setting reflects standard deep MARL architectures, in which the majority of parameters arise from square hidden-to-hidden transformations. We consider a shared actor network used by $A$ agents, where $A$ denotes the number of agents in the parameter-sharing scheme.

Under weight-space masking, each agent maintains $A$ elementwise mask over the shared weight matrix. The total parameter count becomes
\[
P_{\text{weight}}=d^2+Ad^2=(A+1)d^2.
\]

Under spectral masking, the shared SVD-parameterized representation $\mathbf{W}_\theta = \mathbf{U} \, \mathrm{diag}(\mathbf{s}) \, \mathbf{V^\top}$ contributes $2d^2+d$ parameters, and each agent introduces a spectral mask $\mathbf{m}_i \in \mathbb{R}^d$. This yields
\[
P_{\text{spectral}}=2d^2+d+Ad.
\]

The parameter-efficiency gap is therefore
% \[
% \Delta =P_{\text{weight}}-P_{\text{spectral}}=(A+1)d^2-(2d^2+d+Ad)=A(d^2-d)-(d^2+d).
% \]
\begin{align}
\Delta
&= P_{\text{weight}} - P_{\text{spectral}} \\
&= (A+1)d^2 - (2d^2 + d + Ad) \\
&= A(d^2 - d) - (d^2 + d).
\end{align}

Spectral masking is more parameter efficient when $\Delta > 0$, which holds exactly when
\[
A(d-1)>d+1 \Longleftrightarrow A> \frac{d+1}{d-1}.
\]

For typical dimensions $d \in [32, 512]$, the threshold $\frac{d+1}{d-1} \approx 1.01 \sim 1.06$. Therefore, any multi-agent system with at least two agents ($A \geq 2$) satisfies the condition, implying that spectral masking is strictly more parameter-efficient than weight-space masking in all realistic MARL settings.

The comparison above corresponds to the worst-case setting where each agent may individually control all $d$ singular directions. In practice, \textbf{Prism} applies masking only to the non-shared portion of the spectrum, whose size is $(1-\rho)d$. Increasing the shared ratio $\rho$ therefore reduces the agent-specific mask dimension and further strengthens the parameter-efficiency advantage of spectral masking beyond the full-rank bound derived here.

\section{Additional Ablation Studies}
\subsection{Regularization Robustness}

\begin{figure}[H]
    \centering
    \vspace{0.5em}
    
    \begin{subfigure}[t]{0.23\textwidth}
        \centering
        \includegraphics[width=\textwidth]{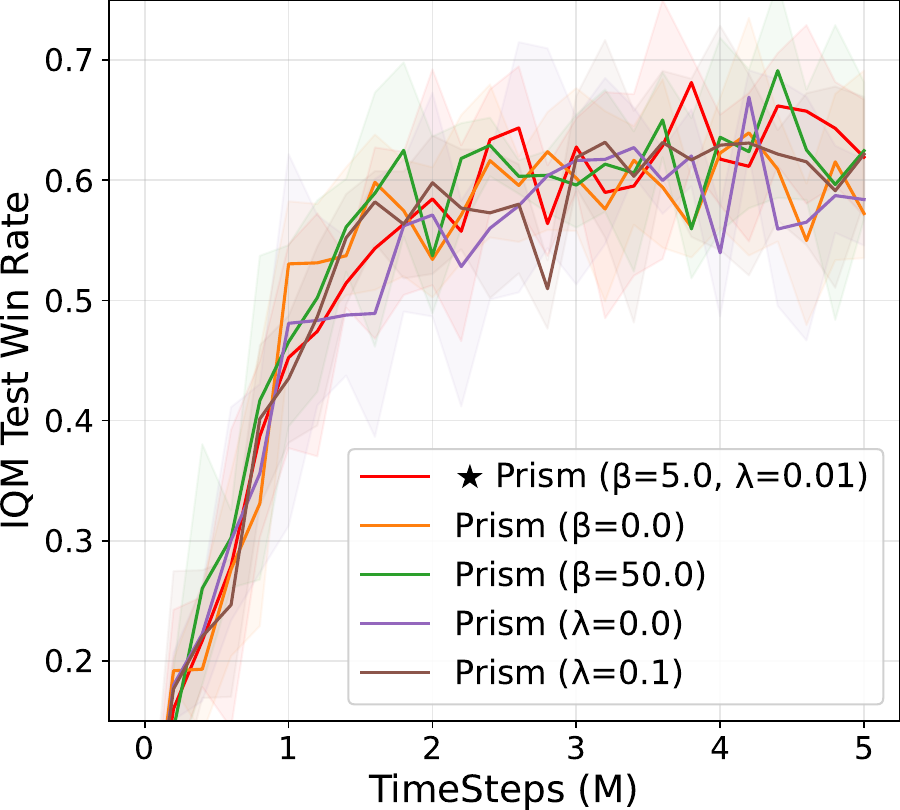}
        \captionsetup{skip=0pt}
        \caption{Terran\_5\_vs\_5}
        \label{fig:further_ablation_reg_smacv2}
    \end{subfigure}%
    \hspace{0.2em}
    \begin{subfigure}[t]{0.23\textwidth}
        \centering
        \includegraphics[width=\textwidth]{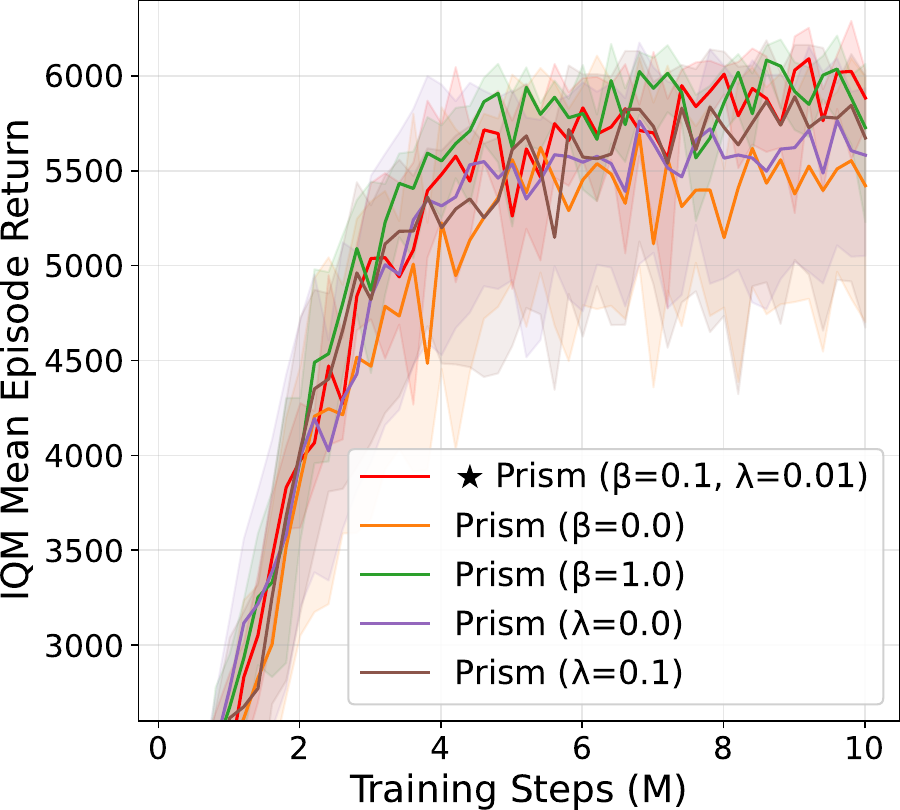}
        \captionsetup{skip=0pt}
        \caption{Ant-v2-4x2}
        \label{fig:further_ablation_reg_mamujoco}
    \end{subfigure}%
    \captionsetup{skip=1pt}
    \caption{Ablation study on the regularization strengths of diversity ($\beta$) and orthogonality ($\lambda$).}
    \label{fig:exp_ablation_reg}
\end{figure}

In addition to the ablations presented in the main paper, we further investigate the robustness of Prism under extreme regularization settings by scaling the diversity and orthogonality coefficients by an order of magnitude (i.e., setting them to 0× and 10× of the default values). This analysis is intended to assess the stability of the proposed regularization terms beyond the moderate ranges considered in Section \ref{sec:exp_ablation}.

Consistent with the main results, the impact of regularization strength is particularly pronounced in heterogeneous environments such as MaMuJoCo. When the diversity regularization is completely removed ($\beta=0$), performance degrades substantially, indicating that explicit encouragement of inter-agent diversity is critical for effective coordination in heterogeneous control tasks.

Similarly, disabling orthogonal regularization ($\lambda = 0$) also leads to a notable performance drop, suggesting that maintaining orthogonality among singular vectors plays an important role in stabilizing training under SVD-parameterized representations.

Overall, these results confirm that both diversity and orthogonal regularization are essential components of Prism, and that removing either term significantly undermines performance, especially in environments where agent heterogeneity is crucial.

\subsection{Spectral Masking}

\begin{figure}[H]
    \centering
    \vspace{0.5em}
    
    \begin{subfigure}[t]{0.23\textwidth}
        \centering
        \includegraphics[width=\textwidth]{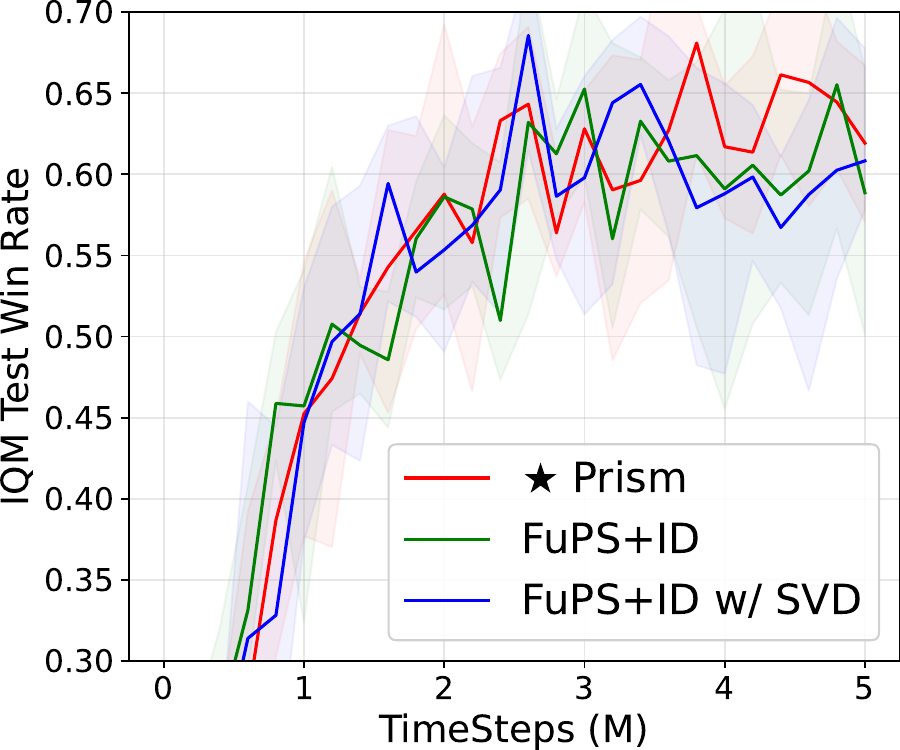}
        \captionsetup{skip=0pt}
        \caption{Terran\_5\_vs\_5}
        \label{fig:ablation_mask_smacv2}
    \end{subfigure}%
    \hspace{0.2em}
    \begin{subfigure}[t]{0.23\textwidth}
        \centering
        \includegraphics[width=\textwidth]{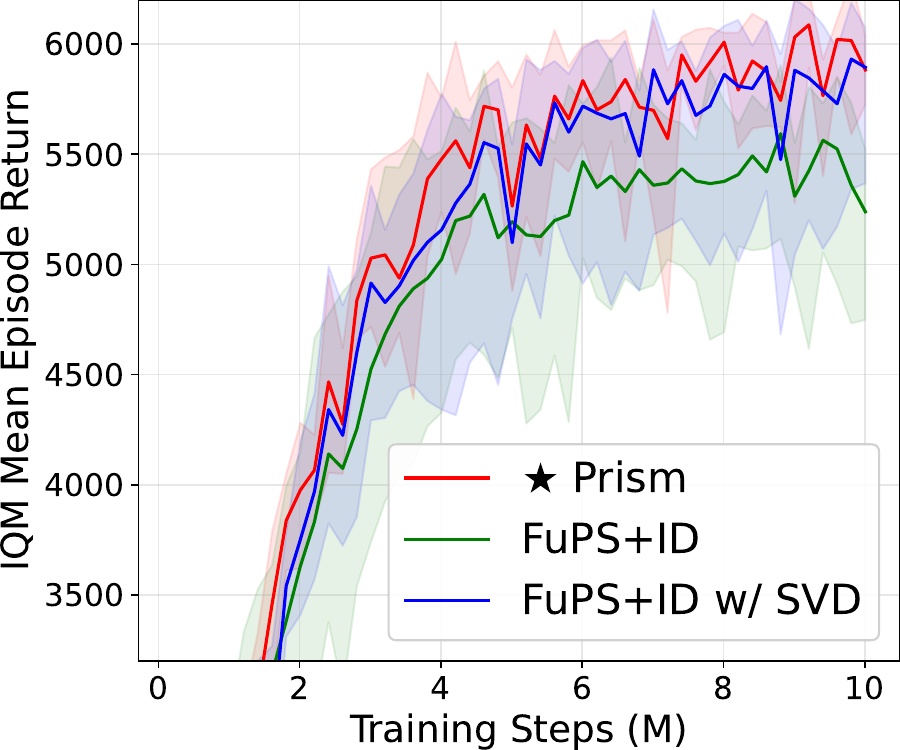}
        \captionsetup{skip=0pt}
        \caption{Ant-v2-4x2}
        \label{fig:ablation_mask_mamujoco}
    \end{subfigure}%
    \captionsetup{skip=1pt}
    \caption{Ablation study on spectral masking.}
    \label{fig:exp_ablation_mask}
\end{figure}

To examine the contribution of spectral parameter sharing to performance gains, we compare three variants: (i) standard parameter sharing with agent identifiers (FuPS+ID), (ii) spectral parameter sharing without diversity regularization (FuPS+ID w/ SVD), and (iii) spectral parameter sharing with learnable spectral masking and diversity regularization (Prism). This comparison isolates the effect of representing shared parameters in the spectral domain and further assesses the benefit of explicit agent differentiation via spectral masking.

The performance difference is particularly evident in the heterogeneous Ant-v2-4×2 task. While spectral parameter sharing already improves performance over conventional parameter sharing, introducing learnable spectral masks with diversity regularization yields additional gains. This result highlights that explicitly allocating agent-specific spectral subspaces is critical for realizing the full benefit of spectral parameter sharing in heterogeneous environments.

\section{Hyperparameters}
\begin{table}[H]
\centering

\label{tab:lab_hyperparameter}
\begin{tabular}{l c}
\hline
\textbf{Common Hyperparameters} & \textbf{Value} \\
\hline
Action selector            & Epsilon-greedy \\
Epsilon start              & 1.0 \\
Epsilon finish             & 0.05 \\
Epsilon anneal time        & $5 \times 10^{4}$ \\
Runner                     & Episode \\
Number of environments     & 1 \\
Replay buffer size         & $5 \times 10^{3}$ \\
Batch size                 & 32 \\
Max steps                  & $3 \times 10^{6}$ \\
Target update interval     & 200 \\
Number of agents           & 3 \\
Number of foods            & 3 \\ 
Hidden dimension           & 64 \\
Mixing embedding dimension & 32 \\
Hypernetwork embedding dim & 64 \\
Learning rate              & 0.0005 \\
Grad norm clip             & 10 \\
Discount factor ($\gamma$) & 0.99 \\
TD Lambda                  & 0.6 \\
Q Lambda                   & False \\
\hline
\end{tabular}
\caption{Common hyperparameters used in LBF.}
\end{table}

\begin{table}[H]
\centering
\label{tab:smacv2_hyperparameter}
\begin{tabular}{l c}
\hline
\textbf{Common Hyperparameters} & \textbf{Value} \\
\hline
Action selector            & Epsilon-greedy \\
Epsilon start              & 1.0 \\
Epsilon finish             & 0.05 \\
Epsilon anneal time        & $1 \times 10^{5}$ \\
Runner                     & Parallel \\
Number of environments     & 4 \\
Replay buffer size         & $5 \times 10^{3}$ \\
Batch size                 & 128 \\
Max steps                  & $5 \times 10^{6}$ \\
Target update interval     & 200 \\
Number of allied agents    & 5 \\
Number of enemy agents     & 5 \\
Number of unit types       & 3 \\
Hidden dimension           & 64 \\
Mixing embedding dimension & 32 \\
Hypernetwork embedding dim & 64 \\
Learning rate              & 0.001 \\
Grad norm clip             & 10 \\
Discount factor ($\gamma$) & 0.99 \\
TD Lambda                  & 0.6 \\
Q Lambda                   & False \\
\hline
\end{tabular}
\caption{Common hyperparameters used in SMACv2.}
\end{table}
\begin{table}[H]
\centering
\label{tab:mamujoco_hyperparameter}
\begin{tabular}{l c}
\hline
\textbf{Common Hyperparameters} & \textbf{Value} \\
\hline
Batch size              & 1000 \\
Replay buffer size      & $1 \times 10^{6}$ \\
Exploration noise       & 0.1 \\
Fixed order             & True \\
Discount factor ($\gamma$) & 0.99 \\
$n$-step return         & 5 \\
Noise clipping          & 0.5 \\
Policy update frequency & 2 \\
Policy noise            & 0.2 \\
Polyak coefficient      & 0.005 \\
Number of critics       & 5 \\
Activation function     & ReLU \\
Final activation        & Tanh \\
Actor learning rate     & $5 \times 10^{-4}$ \\
Critic learning rate    & 0.001 \\
Hidden dimension        & 256 \\
Max steps               & $1 \times 10^{7}$ \\
Training interval       & 50 \\
Warmup steps            & $1 \times 10^{4}$ \\
\hline
\end{tabular}
\caption{Common hyperparameters used in MaMuJoCo.}
\end{table}

\section{Resource Constraint Parameter}
This appendix reports the exact number of parameters used in the resource-constrained experiments presented in Section 5.3. For each method, the network width is adjusted to closely match the target parameter budget, resulting in nearly identical parameter counts across methods.
\begin{table}[H]
\centering
\label{tab:terran_resource_constraint}
\begin{tabular}{c c c c c}
\hline
\textbf{Ratio} & \textbf{NoPS} & \textbf{SNP} & \textbf{Kaleidoscope} & \textbf{Prism} \\
\hline
1.0 & 196{,}535 & 194{,}187 & 197{,}715 & 194{,}997 \\
0.5 & 99{,}935 & 99{,}851 & 99{,}761 & 100{,}553 \\
\hline
\end{tabular}
\caption{The number of parameters in resource constraint SMACv2 (\textit{Protoss\_5\_vs\_5}).}
\end{table}

\begin{table}[H]
\centering
\label{tab:protoss_resource_constraint}
\begin{tabular}{c c c c c}
\hline
\textbf{Ratio} & \textbf{NoPS} & \textbf{SNP} & \textbf{Kaleidoscope} & \textbf{Prism} \\
\hline
1.0 & 199{,}735 & 195{,}667 & 195{,}137 & 198{,}382 \\
0.5 & 102{,}135 & 100{,}891 & 102{,}561 & 101{,}453 \\
\hline
\end{tabular}
\caption{The number of parameters in resource constraint SMACv2 (\textit{Terran\_5\_vs\_5}).}
\end{table}
\begin{table}[H]
\centering
\label{tab:ant_resource_constraint}
\begin{tabular}{c c c c c}
\hline
\textbf{Ratio} & \textbf{NoPS} & \textbf{SNP} & \textbf{Kaleidoscope} & \textbf{Prism} \\
\hline
1.0 & 379{,}912 & 380{,}008 & 377{,}624 & 380{,}536 \\
0.5 & 190{,}184 & 191{,}150 & 188{,}786 & 190{,}996 \\
\hline
\end{tabular}
\caption{The number of parameters in resource constraint MaMuJoCo (\textit{Ant-v2-4x2}).}
\end{table}

\begin{table}[H]
\centering
\label{tab:walker_resource_constraint}
\begin{tabular}{c c c c c}
\hline
\textbf{Ratio} & \textbf{NoPS} & \textbf{SNP} & \textbf{Kaleidoscope} & \textbf{Prism} \\
\hline
1.0 & 142{,}342 & 141{,}969 & 142{,}143 & 143{,}415 \\
0.5 & 71{,}206 & 71{,}129 & 70{,}719 & 70{,}615 \\
\hline
\end{tabular}
\caption{The number of parameters in resource constraint MaMuJoCo (\textit{Walker2D-v2-2x3}).}
\end{table}

% \section*{Acknowledgments}
% \input{sections/acknowledgements}

%%%%%%%%%%%%%%%%%%%%%%%%%%%%%%%%%%%%%%%%%%%%%%%%%%%%%%%%%%%%%%%%%%%%%%%%

\bibliographystyle{named}
\bibliography{ijcai26}

\end{document}